\documentclass[journal]{IEEEtran}

\usepackage{enumerate}

\usepackage{tabularx}
\usepackage{makecell}
\usepackage{bbm}

\usepackage[table,xcdraw]{xcolor}
\usepackage{threeparttable}
\usepackage{algpseudocode}
\usepackage{setspace}

\usepackage{stfloats}
\usepackage{amsfonts}
\usepackage{amssymb}
\usepackage{amsthm}
\usepackage{cite}
\usepackage[cmex10]{amsmath}
\usepackage{float}
\usepackage{color}
\usepackage{stfloats,fancyhdr}
\usepackage{amsmath,bm}
\usepackage{algorithm}
\usepackage{multirow}
\usepackage{changepage}
\usepackage[normalem]{ulem}
\usepackage{balance}

\ifCLASSINFOpdf
\usepackage[pdftex]{graphicx}
\DeclareGraphicsExtensions{.pdf,.jpeg,.png}
\else
\usepackage[dvips]{graphicx}
\DeclareGraphicsExtensions{.eps}
\fi

\usepackage{subfigure}
\usepackage{fancybox,dashbox}
\usepackage{authblk}
\usepackage{tabularx}

\usepackage{mathtools}
\mathtoolsset{showonlyrefs=true}

\begin{document}

\title{Communication-Control Codesign for Large-Scale Wireless Networked Control Systems}

\author{Gaoyang Pang, Wanchun Liu*,~\IEEEmembership{Member,~IEEE,}
Dusit Niyato,~\IEEEmembership{Fellow,~IEEE,} 
Branka Vucetic,~\IEEEmembership{Life Fellow,~IEEE,}
Yonghui Li,~\IEEEmembership{Fellow,~IEEE} 
\vspace{-0.5cm}
\thanks{The work of W. Liu was supported by the Australian Research Council’s Discovery Early Career Researcher Award (DECRA) Project DE230100016. \textit{(Corresponding author: W. Liu.)}}
\thanks{G. Pang, W. Liu, B. Vucetic, and Y. Li are with the School of Electrical and Computer Engineering, The University of Sydney, Sydney, NSW 2006, Australia (e-mail: gaoyang.pang@sydney.edu.au; wanchun.liu@sydney.edu.au; branka.vucetic@sydney.edu.au; yonghui.li@sydney.edu.au).}
\thanks{D. Niyato is with the College of Computing and Data Science, Nanyang Technological University, Singapore 639798, (e-mail: dniyato@ntu.edu.sg).}
} 

\maketitle

\begin{abstract}
Wireless Networked Control Systems (WNCSs) are essential to Industry 4.0, enabling flexible control in applications, such as drone swarms and autonomous robots. The interdependence between communication and control requires integrated design, but traditional methods treat them separately, leading to inefficiencies. Current codesign approaches often rely on simplified models, focusing on single-loop or independent multi-loop systems. However, large-scale WNCSs face unique challenges, including coupled control loops, time-correlated wireless channels, trade-offs between sensing and control transmissions, and significant computational complexity. To address these challenges, we propose a practical WNCS model that captures correlated dynamics among multiple control loops with spatially distributed sensors and actuators sharing limited wireless resources over multi-state Markov block-fading channels. We formulate the codesign problem as a sequential decision-making task that jointly optimizes scheduling and control inputs across estimation, control, and communication domains. To solve this problem, we develop a Deep Reinforcement Learning (DRL) algorithm that efficiently handles the hybrid action space, captures communication-control correlations, and ensures robust training despite sparse cross-domain variables and floating control inputs. Extensive simulations show that the proposed DRL approach outperforms benchmarks and solves the large-scale WNCS codesign problem, providing a scalable solution for industrial automation.
\end{abstract}

\begin{IEEEkeywords}
Wireless networked control, Industry 4.0, transmission scheduling, and deep reinforcement learning.
\end{IEEEkeywords}

\section{Introduction} \label{sec:intro}
Wireless Networked Control Systems (WNCSs) serve as foundational models for driving the next generation of cutting-edge technologies in Industry 4.0, enabling advanced applications such as drone swarms, autonomous vehicles, and mobile robots \cite{jin2023cloud,honghao2024cloud,bhimavarapu2022unobtrusive}. By eliminating the physical constraints of wired communication, WNCSs offer unprecedented flexibility and scalability, supporting distributed sensing and control across large-scale and complex industrial environments. However, the transition to wireless communication introduces fundamental challenges, particularly in maintaining robust control performance amidst uncertainties such as delays, packet loss, and limited wireless resources \cite{CoDesign}. These wireless communication challenges pose a significant threat to system stability and reliability, necessitating a fundamental rethinking of traditional control and communication strategies \cite{CoDesign}.

Communication-control codesign in WNCSs addresses these challenges through a holistic design approach that simultaneously optimizes control algorithms and communication policies. Recognizing the intrinsic interdependence between control and communication ensures that control algorithms compensate for uncertainties in wireless communication while communication strategies adapt to the stringent performance requirements of control tasks. This correlation is critical, as control actions directly influence the timing and priority of data transmission, while the state of the wireless communication network impacts the effectiveness of control signals and sensor feedback. Without an integrated codesign approach, independently optimizing these domains leads to suboptimal system performance, especially in high-stakes applications such as drone swarms and robotics, where precise coordination and timely data exchange are essential for success.

\subsection{Related Work} \label{sec:RelatedWork}
\subsubsection{Control-aware communication design for WNCSs} In this approach, communication policies are optimized under the assumption that control algorithms are fixed, with the goal of improving overall control system performance. Significant research has been dedicated to this design strategy in areas such as power allocation \cite{Gatsis2015Control,knorn2017optimal,Chang2019Power}, transmission scheduling \cite{lu2023jointly,leong2017event,Ji2022Edge,Ji2023Edge,Tzoumas2021LQG,Wang2023TII,Girgis2021TCOMM}, and data rate and packet-length adaptation \cite{Wang2021Control}. However, this method is suboptimal for complex WNCSs because it neglects to account for the design of the control policy. In practice, WNCSs often exhibit tightly coupled dynamics between communication and control, making the two policies highly interdependent \cite{Ji2023Couple}. This interdependence emphasizes the importance of a codesign approach, where communication and control are jointly optimized. Such an approach addresses the limitations of traditional control-aware communication design and reduces its conservatism \cite{lv2022impacts,lyu2024latency}.

\subsubsection{Conventional communication-control codesign methods}
Conventional codesign approaches often decompose the problem into separate subproblems, solving the communication and control policies independently \cite{Demirel2014TAC,Cao2023PAoL}. In this framework, the control policy typically takes the form of linear feedback control (LFC), where the control input is a linear combination of the estimated plant state, derived based on the statistical effects of the communication policy. Once an optimal control policy is determined, the optimal communication scheduling is selected from the policy space to enhance control performance \cite{di2015co,Lu2023TII,9945199,Chen2020TII}. Although communication and control are optimized iteratively, this step-by-step optimization can lead to suboptimal performance, as communication decisions are made without considering their future impact on control performance. Moreover, recalculating the control policy at each time slot, based on updated communication policies, introduces inefficiencies and instability in dynamic environments \cite{Chen2020TII}. The failure to fully integrate communication and control policies results in missing opportunities to leverage their interdependencies, which limits overall system optimization.

Another class of methods employs model predictive control (MPC), where the optimization problem is solved iteratively, producing a finite sequence of communication actions and control inputs over a prediction horizon \cite{wildhagen2020scheduling,9210732}. MPC accounts for the future impact of current decisions and jointly optimizes both policies. However, finding the best scheduling policy over long prediction horizons often requires exhaustive or heuristic searches, which can be computationally intensive \cite{Peters2016Codesign}. While recent research has explored approximations and simplifications to reduce computational complexity and improve the applicability of MPC in WNCSs \cite{8796135,yao2020contention,cui2020co,bahraini2022optimal}, these strategies may compromise the accuracy and effectiveness of the policies. The computational burden of MPC remains a significant limitation, particularly for large-scale systems or those requiring real-time optimization, hindering its scalability and practical deployment.

\subsubsection{Codesign approaches with deep reinforcement learning}
To overcome the limitations of conventional codesign methods, deep reinforcement learning (DRL) has emerged as a promising alternative for jointly optimizing communication and control policies \cite{redder2019deep}. DRL provides a unified framework that learns optimal policies by interacting with the environment, effectively capturing the interdependencies between communication and control systems. Unlike conventional approaches that focus on short-term gains, DRL leverages cumulative costs, allowing for decisions that consider the long-term impact of current actions on future states and control performance. By employing deep neural networks (DNNs), DRL also addresses computational inefficiencies in traditional methods, enabling online execution with reduced computational overhead.

In scenarios with \emph{perfect channels}, Baumann et al. \cite{8619335} proposed a DRL algorithm to maximize control performance in single-loop WNCS with limited transmissions. Funk et al. \cite{funk2021learning} enhanced this by introducing a hierarchical DRL framework, where the algorithm first decides whether to transmit and then computes control inputs. Kesper et al. \cite{kesper2023toward} and Shibata et al. \cite{9561274,shibata2023deep} extended these techniques to multi-loop WNCS. Similar DRL methods for single-loop WNCS can be found in \cite{wan2023model,wan2023integrated,wang2023deep}.

For scenarios with \emph{imperfect channels}, Termehchi et al. \cite{Termehchi} developed a hierarchical DRL algorithm that jointly optimizes actuator scheduling, control policies, channel allocation, and power allocation in a single-loop WNCS. Zhao et al. \cite{Zhao2023IoTJ} introduced a DRL approach to jointly learn estimation, communication, and control policies in a single-loop WNCS, while Lima et al. \cite{lima2022model} designed a DRL algorithm that jointly optimizes power allocation and control in a multi-loop WNCS.

These works demonstrate the effectiveness of DRL in handling the complexities of WNCSs, where communication and control systems are tightly coupled. However, significant challenges remain, especially in scaling these approaches for large, practical multi-loop WNCSs with limited wireless resources and dynamic environments.

\subsection{Motivations}
\subsubsection{DRL-based codesign for large-scale systems}
Existing DRL methods are largely constrained to small-scale WNCSs due to oversimplified communication models \cite{8619335,Zhao2023IoTJ,funk2021learning,wan2023model,wan2023integrated,wang2023deep,Termehchi}. Scaling these methods to more complex systems presents significant challenges that limit their effectiveness. One major issue is the hybrid action space that arises from the need to handle both discrete communication actions (e.g., channel allocation) and continuous control inputs (e.g., actuator commands). As the system grows in size, the hybrid action space expands exponentially, making it increasingly difficult for DRL algorithms to explore and learn optimal policies effectively. This leads to a dramatic increase in computational complexity and slows down convergence, posing a fundamental limitation on the scalability of DRL-based solutions.

Furthermore, the complexity is compounded by the presence of cross-domain variables—where inputs and outputs from communication, control, and estimation domains interact. These variables often have widely differing scales, such as discrete, finite channel assignments versus continuous, potentially infinite control actions. This mismatch in scale can result in issues such as gradient explosion  \cite{lei2016layer}, leading to unstable training processes and making it harder for DRL models to converge on effective solutions. The intricate coupling of these domains further complicates the problem, as decisions made in one domain (e.g., communication) have cascading effects on the others (e.g., control and estimation), adding another layer of complexity to the training process.
As a result, current DRL methods fail to efficiently handle large-scale systems with hybrid action spaces and cross-domain interactions, severely limiting their scalability and applicability in real-world, large-scale WNCSs.

\subsubsection{Model practicality issues of WNCSs}
\begin{figure}[t]
    \centering
    \includegraphics[width=3.5in]{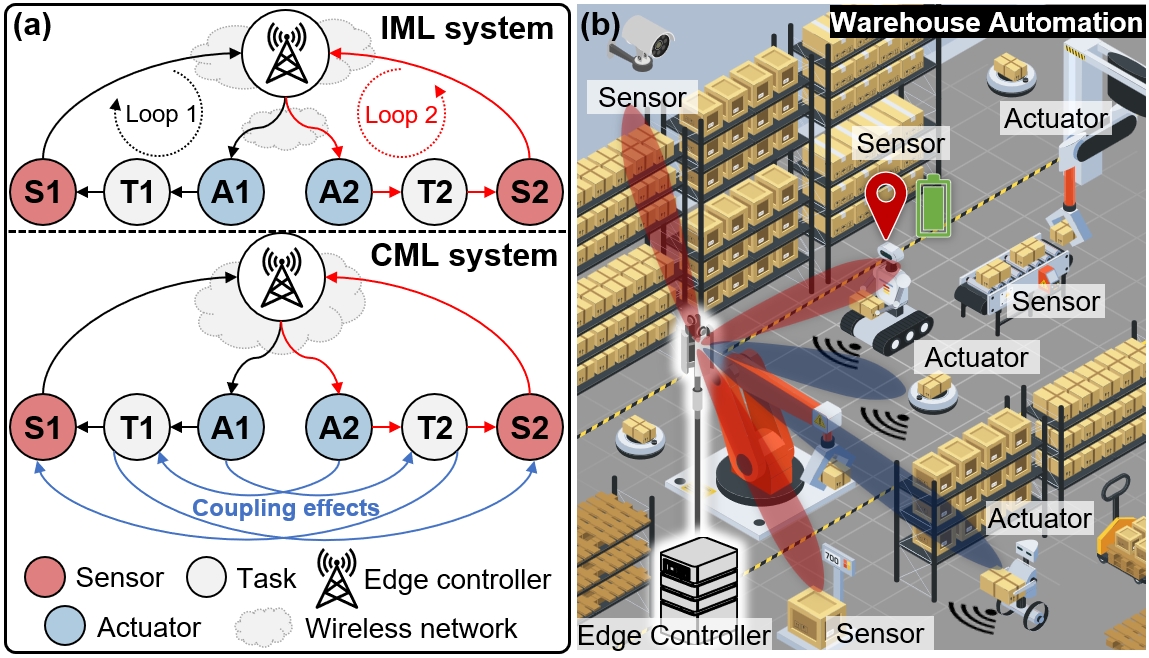}
    \vspace{-0.8cm}
    \caption{Illustration of multi-loop WNCSs: (a) Network topology of IML and CML systems. (b) CML WNCSs in warehouse automation.} 
    \label{fig:CML_WNCS}
    \vspace{0.0cm}
\end{figure}

The practicality of existing multi-loop WNCS models faces two significant issues: the lack of coupling effects in plant state dynamics and oversimplified communication models.

\textbf{Lack of Coupling Effect in Control.} Most existing works focus on single-loop (SL) systems \cite{Cao2023PAoL,Demirel2014TAC,di2015co,10233620,8796135,cui2020co,8619335,funk2021learning,wan2023model,wan2023integrated,wang2023deep,Termehchi,Zhao2023IoTJ} or independent multi-loop (IML) systems \cite{Lu2023TII,Chen2020TII,9945199,yao2020contention,kesper2023toward,9561274,shibata2023deep,redder2019deep,lima2022model}. In WNCSs, a control loop refers to the closed feedback loop between sensors, controllers, and actuators, where the system state is continuously monitored by sensors, decisions are made by controllers, and corrective actions are taken by actuators.
In IML systems (see Fig.~\ref{fig:CML_WNCS}(a)), control loops operate independently, and controllers are designed separately for each loop, focusing solely on local performance in each loop only. In this case, plants (physical systems being controlled) are independent, meaning the state of one plant does not directly influence the state or control strategy of another plant. Each control loop is concerned with optimizing its own system's performance without considering interactions or dependencies with other control loops.

In contrast, coupled multi-loop (CML) systems (see Fig.~\ref{fig:CML_WNCS}(a)) involve interdependent control loops, where the states of different plants are correlated. Sensor measurements from one plant may be influenced by the states of others, and control actions taken for one plant can affect the dynamics of others. For instance, an actuator’s action may alter the behavior of other plants due to shared environments or constraints. An example is shown in Fig.~\ref{fig:CML_WNCS}(b), where robots navigate different sections to retrieve and deliver goods in a warehouse. The edge controller coordinates robot actions by receiving sensor data (e.g., position and load) via uplink transmissions and sending control commands (e.g., speed and braking) via downlink. These control loops are interdependent; for example, if one robot slows down to avoid a collision, it may delay others reliant on timely item handovers. Such interdependencies are critical for coordinated and efficient control in CML systems, which are rarely considered in the literature. 

\textbf{Lack of Coupling Effect in Communications.}
Existing works on multi-loop WNCSs rarely consider joint uplink (sensor-to-controller) and downlink (controller-to-actuator) communication scheduling with shared wireless resources. In most studies, communication scheduling is treated separately for uplink and downlink transmissions, often focusing on a specific type of communication while neglecting the resource contention between the two.

However, in practical WNCSs, both uplink and downlink communications compete for the same limited wireless resources, making joint scheduling essential. The state estimation accuracy relies on timely sensor data (uplink), while control performance depends on the delivery of control commands (downlink). Failing to jointly schedule these transmissions can lead to inefficiencies, such as over-allocating resources to sensor updates when control commands are more urgent, or vice versa. For example, in a warehouse with multiple robots, allocating too many resources for sensors to collect position data might delay critical control commands needed to prevent a collision. Thus, in CML WNCSs, the communication and control processes are deeply intertwined, and joint uplink-downlink scheduling is crucial to balance resource allocation and ensure optimal system performance.

\subsection{Contributions}
In this work, we tackle the above challenges. The novel contributions are summarized below.

\subsubsection{\textbf{Practical model of WNCSs}}
We propose a practical model for CML WNCSs that addresses the real-world complexities of distributed sensing and actuation, with sensors and actuators sharing limited wireless resources for both uplink and downlink transmissions. Unlike existing models that often rely on oversimplified assumptions, our approach captures the interdependencies between plant dynamics and communication, recognizing that the overall system performance hinges on both communication and control policies. By incorporating short-packet communications and modeling wireless channels using multi-state block Markov fading \cite{Sadeghi2008FSMC}, the model reflects the time-correlated nature of wireless channels prevalent in industrial settings. 
Additionally, to account for the limited wireless resources and varying channel quality, we introduce state estimation strategies that estimate the overall system state based on partially observed sensor data when not all sensor packets are successfully transmitted. This comprehensive and practical model offers a robust framework for solving the communication-control codesign problem, enabling its application in large-scale WNCS environments where real-world constraints are crucial.

\subsubsection{\textbf{Novel problem formulation}}
We establish a novel Markov Decision Process (MDP) formulation for the communication-control codesign problem of the WNCS. The MDP integrates a comprehensive state representation that includes estimated plant states, estimation quality states, channel quality states, and control quality states. This enriched state space uniquely captures the complex interdependencies between the control system and the wireless network, enabling more context-aware decision-making. The action space includes both the dynamic allocation of frequency channels and the generation of control inputs, addressing communication and control tasks simultaneously under resource constraints. By unifying these elements within the MDP framework, our approach advances communication-control codesign strategies, improving control performance in practical CML WNCS environments.
The codesign MDP introduces several complex challenges: the management of a large and hybrid action space (both discrete and continuous), the need to capture the correlation between communication and control policies, and handling cross-domain variables, all while maintaining training efficiency.

\subsubsection{\textbf{Advanced DRL algorithm}}
We propose an innovative Graph-enhanced Cascaded Actor DRL (GCA-DRL) algorithm to address the challenges of communication-control codesign. The dual-branch architecture encodes scheduling and control policies separately. The scheduling branch generates a continuous weight matrix for device-to-channel assignment, mapped to discrete actions using a graph-based embedding that leverages domain-specific communication knowledge for efficient exploration. This output is cascaded into the control branch, allowing real-time adaptation of control decisions based on communication resource allocation and effectively capturing the interdependencies between policies.
To manage cross-domain variables, layer normalization is applied after each hidden layer, stabilizing training and balancing learning across different variable scales. A partial policy gradient update mechanism further enhances efficiency by focusing only on gradients from effective control inputs, filtering out noise from unscheduled actuators. These innovations improve training stability, convergence, and performance, with extensive numerical results demonstrating that the proposed algorithm outperforms all benchmark policies, making it highly effective for optimizing communication-control codesign in large-scale WNCSs.

\textbf{Outlines.} The rest of this article is organized as follows. The proposed system model of the WNCS is described in Section~\ref{sec:sys}. The codesign problem formulation is presented in Section~\ref{sec:Formulation}. The advanced DRL algorithm to solve the codesign problem is detailed in Section~\ref{sec:DRL}. Numerical simulation is demonstrated and discussed in Section~\ref{sec:simulation}, followed by a conclusion in Section~\ref{sec:conclusion}.

\textbf{Notations.} Matrices and vectors are denoted by capital and lowercase upright bold letters, e.g., $\mathbf{A}$ and $\mathbf{a}$, respectively. $|\mathbf{v}|$ is the Euclidean norm of vector $\mathbf{v}$. $\mathbb{E}\left[\cdot\right]$ is the expectation operator. $\left[\mathbf{A}\right]_{i,j}$ denotes the element at $i$-th row and $j$-th column of a matrix $\mathbf{A}$. $\rho_k\left(\mathbf{A}\right)$ represents the $k$-th eigenvalues of the matrix $\mathbf{A}$. $(\cdot)^\top$ is the vector or matrix transpose operator. $\operatorname{Tr}(\cdot)$ is the matrix trace operator. $\operatorname{Pr}[\cdot]$ is the probability operator. $\mathbb{R}$ denotes the set of real numbers. 

\section{Proposed Multi-loop WNCS Model} \label{sec:sys}
\begin{figure}[t]
    \centering
    \includegraphics[width=3.3in]{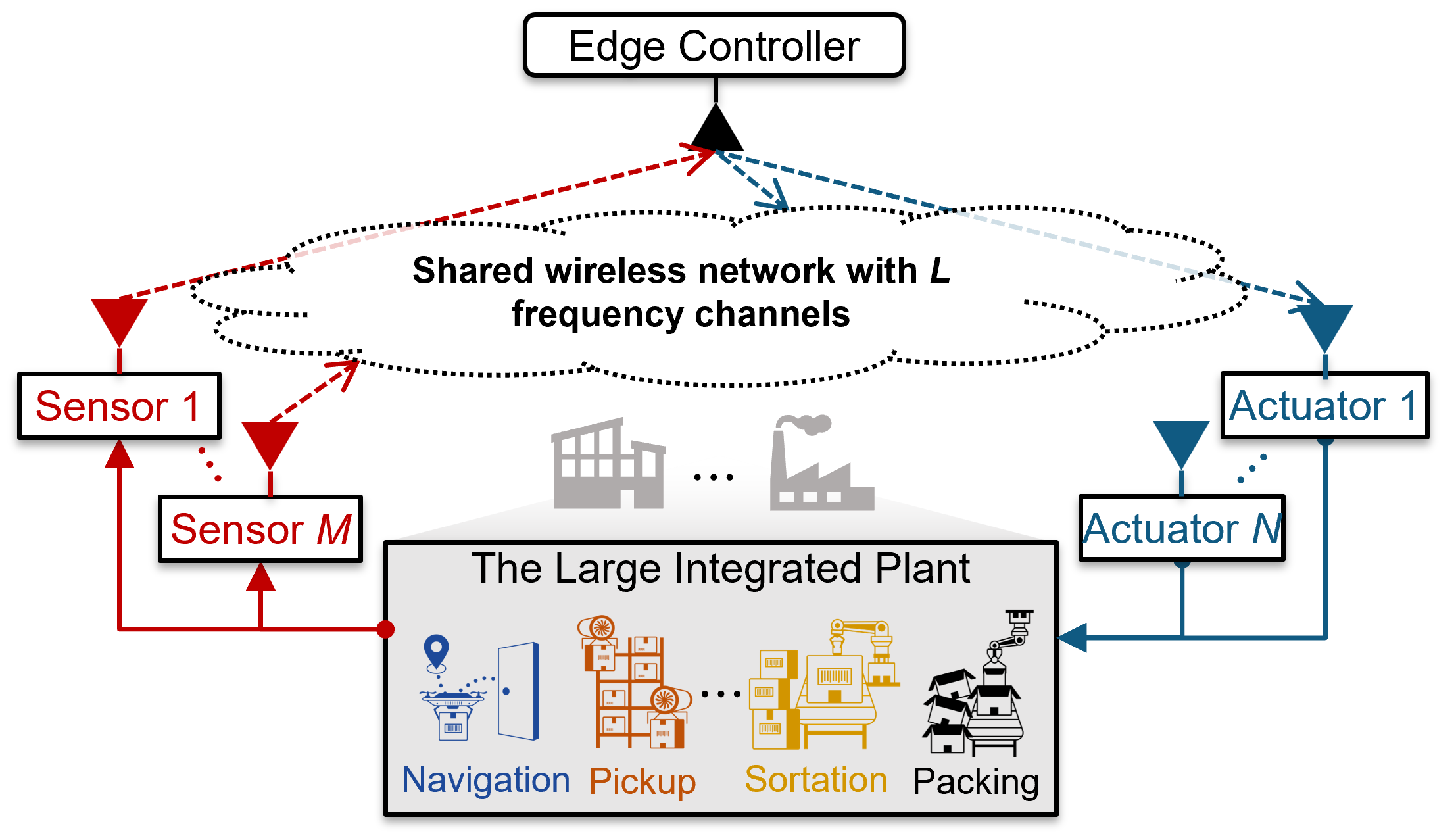}
    \vspace{-0.4cm}
    \caption{Illustration of the core components a CML WNCS.} 
    \label{fig:CML_Components}
    \vspace{0.0cm}
\end{figure}

We consider a multi-loop WNCS consisting of an intelligent edge controller, $M$ spatially distributed sensors, and $N$ actuators within a large integrated plant, as shown in Fig.~\ref{fig:CML_Components}. The large integrated plant is modeled as a collection of interdependent control tasks spanning multiple plants. Sensors transmit state measurements to the edge controller, while actuators execute the control commands received from the edge controller. All devices are equipped with a single antenna for communication. Communication links for all devices are established over $L$ wireless frequency channels (i.e., subcarriers), where $1 \leq L \leq N + M$. The index sets of channels, sensors, and actuators are denoted as $\mathcal{L}\!\!\triangleq\!\!\{1,2,\dots,L\}$, $\mathcal{M}\!\!\triangleq\!\!\{1,2,\dots,M\}$, and $\mathcal{N}\!\!\triangleq\!\!\{1,2,\dots,N\}$. The edge controller implements an estimator to track the plant states and employs a dynamic policy for joint channel allocation and control.

\subsection{Plant Dynamics}
The plant dynamics is modeled as a linear discrete-time system with a sampling time period of $T_s$:
\begin{subequations} \label{LTI}
\begin{align}
    \mathbf{x}(t+1) &=\mathbf{A}\mathbf{x}(t)+\mathbf{B}\mathbf{u}^{\mathrm{rx}}(t)+\mathbf{w}(t),\label{LTI,a}\\
    \mathbf{y}^{\mathrm{tx}}(t) &=\mathbf{C}\mathbf{x}(t) + \mathbf{v}(t),\label{LTI,b}\\
    \mathbf{u}^{\mathrm{rx}}(t) &=\mathbf{\Lambda}(t)\mathbf{u}^{\mathrm{tx}}(t),\label{LTI,c}\\
    \mathbf{y}^{\mathrm{rx}}(t) &=\mathbf{\Psi}(t)\mathbf{y}^{\mathrm{tx}}(t),\label{LTI,d}
\end{align}
\end{subequations}
where $\mathbf{A}\!\!\in\!\!\mathbb{R}^{K\times K}$, $\mathbf{B}\!\!\in\!\!\mathbb{R}^{K\times N}$, $\mathbf{C}\!\!\in\!\!\mathbb{R}^{K\times M}$ are system matrices.

\textbf{Markov plant state dynamics \eqref{LTI,a}.} The plant state is modeled as a multi-dimensional vector $\mathbf{x}(t) \in \mathbb{R}^{K}$, where $K$ represents the number of state dimensions. The evolution of the plant state over time is influenced by two factors: the control input $\mathbf{u}^{\mathrm{rx}}(t)$ received by the plant and a process noise vector $\mathbf{w}(t) \in \mathbb{R}^{K}$. The noise vector $\mathbf{w}(t)$ is assumed to be an independent and identically distributed (i.i.d.) zero-mean Gaussian process with a covariance matrix $\mathbf{W}$. The next state $\mathbf{x}(t+1)$ depends on the current state $\mathbf{x}(t)$, the control input $\mathbf{u}^{\mathrm{rx}}(t)$, and the plant noise $\mathbf{w}(t)$. The set of all state dimensions is denoted as $\mathcal{K} \triangleq \{1, 2, \dots, K\}$. We assume that each state dimension is unstable, i.e., $\rho_k(\mathbf{A}) > 1, \forall k \in \mathcal{K}$, meaning that without control inputs, the state vector $\mathbf{x}(t)$ will grow unbounded over time, i.e., $\mathbf{u}^{\mathrm{rx}}(t) = 0, \forall t$.

\textbf{Transmitted and received sensor measurements \eqref{LTI,b} and \eqref{LTI,d}.} The plant's sensor measurements are given by $\mathbf{y}^{\mathrm{tx}}(t)$, where $\mathbf{C} \in \mathbb{R}^{M \times K}$ is the measurement matrix that maps the true plant state $\mathbf{x}(t)$ to the observed sensor outputs. Since sensors may only capture part of the full system state, $\mathbf{C}$ reflects this partial observability. The noise vector $\mathbf{v}(t) \in \mathbb{R}^{M}$, modeled as an i.i.d. zero-mean Gaussian process with covariance matrix $\mathbf{V}$, further distorts the accuracy of the measurements. Each measurement corresponds to an individual sensor. Due to potential packet loss and limited wireless resources, the edge controller may not receive all sensor measurements. The received measurements at the controller are denoted as $\mathbf{y}^{\mathrm{rx}}(t) \in \mathbb{R}^{M}$, where $\mathbf{\Psi}(t) \in {0,1}^{M \times M}$ is a binary diagonal matrix that indicates successful packet transmissions. Specifically, $\phi_{m}(t) \triangleq [\mathbf{\Psi}(t)]_{m,m} = 1$ means that the $m$-th sensor's measurement has been successfully received by the edge controller.

\textbf{Transmitted and received control inputs \eqref{LTI,c}.} The plant control input $\mathbf{u}^{\mathrm{rx}}(t) \in \mathbb{R}^{N}$ consists of $N$ dimensions, with each dimension corresponding to a control signal for one of the $N$ actuators. The control signals transmitted from the edge controller are represented as $\mathbf{u}^{\mathrm{tx}}(t) \in \mathbb{R}^{N}$. Due to unreliable wireless communication, not all control signals are successfully received by the actuators. This is captured by the binary matrix $\mathbf{\Lambda}(t) \in \{0,1\}^{N \times N}$, where $\lambda_{n}(t) \triangleq [\mathbf{\Lambda}(t)]_{n,n} = 1$ indicates successful transmission to the $n$-th actuator. If a control signal is not received (i.e., $\lambda_n(t) = 0$), the actuator follows a zero-input policy.

This assumption, where each sensor measures a single dimension and each actuator is associated with one control signal, is commonly used in industrial automation \cite{Peters2016Codesign,Wang2023TII,Wang2021Control}. It simplifies the system by avoiding redundant measurements and overlapping feedback, ensuring more precise control. Although multi-dimensional sensors and actuators exist, focusing on single-dimensional devices makes the codesign problem more tractable without sacrificing generality.

\subsection{Wireless Communications}
Communication channels are commonly time-correlated and are quantized to finite discrete states to reflect dynamic channel properties of complex industrial indoor environments with many mobile machines, vehicles, and robots. Thus, we consider finite-state time-homogeneous Markov block-fading channels \cite{Sadeghi2008FSMC}. The channel states remain constant during a time slot of $T_s$ and changes slot-by-slot. The edge controller has the knowledge of channel state information achieved by standard channel estimation and feedback techniques. 

Let $\mathbf{g}_{m}^{s}(t) \triangleq(g_{m,1}^{s}(t), g_{m,2}^{s}(t), \ldots, g_{m,L}^{s})^\top, \forall m \in \mathcal{M}$ and $\mathbf{g}_{n}^{c}(t) \triangleq(g_{n,1}^{c}(t), g_{n,2}^{c}(t), \ldots, g_{n,L}^{c})^\top, \forall n \in \mathcal{N}$ represent the channel power gain of sensor $m$ and actuator $n$ at $L$ channels, respectively. Each channel power gain has $H$ states, i.e., $ \mathcal{G} \triangleq\left\{h_{1}, h_{2}, \ldots, h_{H}\right\}$ and is modelled as a multi-state Markov chain. Since devices are dislocated and have different radio propagation environments, we assume that the Markov channel states are independent. The transition matrices of channel power gain for sensors and actuators at $L$ channels are denoted as $\mathbf{M}_{m,l}^{s} \in \mathbb{R}^{H \times H}, \forall m \in \mathcal{M}$ and $\mathbf{M}_{n,l}^{c} \in \mathbb{R}^{H \times H}, \forall n \in \mathcal{N}$, respectively. The overall channel state information (CSI) is $\mathbf{G}(t)\triangleq\left[\mathbf{g}_{1}^{s}(t),\dots,\mathbf{g}_{M}^{s}(t),\mathbf{g}_{1}^{c},\dots,\mathbf{g}_{N}^{c}(t)\right] \in \mathcal{G}^{L \times (M+N)} \triangleq \{\tilde{\mathbf{G}}_1,\tilde{\mathbf{G}}_2,\dots,\tilde{\mathbf{G}}_{L \times (M+N)}\}$. Let $\mathbf{M} \in \mathbb{R}^{H^{L \times (M+N)} \times H^{L \times (M+N)}}$ denote the overall channel state transition matrix 
\begin{equation}\label{ChannelTransition}
\left[\mathbf{M}\right]_{i, j} \triangleq p_{i, j}^{h}=\operatorname{Pr}\left[\mathbf{G}(t+1)=\tilde{\mathbf{G}}_{j} \mid \mathbf{G}(t)=\tilde{\mathbf{G}}_{i}\right],
\end{equation}
which can be calculated by the Kronecker product of $\{\mathbf{M}_{m,l}^{s}\}$ and $\{\mathbf{M}_{n,l}^{c}\}$. We note that the channel state transition matrix is unknown to the edge controller because the estimation of a multi-dimensional Markov chain model is computationally intensive \cite{He2017FSMC}.

We adopt the short-packet communications for uplink and downlink transmissions \cite{Liu2021Polyanskiy}. In short-packet communications, the decoding failure probability of a packet depends on the packet length, the data bits, and the signal-to-noise ratio (SNR). Thus, the decoding failure probability of a packet can be approximated as \cite{Liu2021Polyanskiy}
\begin{equation}\label{BLER}
\varepsilon\left(\gamma\right) \approx \mathcal{Q}\left(\frac{\mathcal{C}\left(\gamma\right)-\frac{b}{l}}{\sqrt{\frac{\mathcal{V}\left(\gamma\right)}{l}}}\right),
\end{equation}
where $\mathcal{Q}(x)=\frac{1}{\sqrt{2\pi}}\int_{x}^{\infty}{e^{-\frac{t^2}{2}}\mathrm{d}t}$ is the Gaussian Q-function; $l$ is the sensor packet length in symbols; $b$ is the number of data bits; $\gamma$ is the SNR of a packet; $\mathcal{C}(\gamma)=\log_2{(1+\gamma)}$ is the Shannon capacity; $\mathcal{V}(\gamma_n)=(1-(1+\gamma)^{-2})(\log_2{e})^2$ is the channel dispersion.

Each device (either a sensor or an actuator) takes at most one channel for transmission, and each channel is allocated to a single device. In this regard, sensing and control are performed in an off-band full-duplex mode, where a sensing packet and a control packet can be scheduled at different frequency channels at the same time slot.
We denote the channel allocation for sensor $m$ at actuator $n$ at the $L$ channels as binary vectors $\mathbf{d}_{m}^{s}(t)\triangleq(d_{m,1}^{s}(t),\ldots,d_{m,L}^{s}(t))^\top\in\left\{0,1\right\}^L$ and $\mathbf{d}_{n}^{c}(t)\triangleq(d_{n,1}^{c}(t),\ldots,d_{n,L}^{c}(t))^\top\in\left\{0,1\right\}^L$, respectively. 
$d_{m,l}^{s}(t) = 1$ or $d_{n,l}^{c}(t) = 1$ denotes that channel $l$ is allocated to sensor $m$ or actuator $n$, respectively. Thus, the overall channel allocation for all devices is $\mathbf{D}(t)\triangleq [\mathbf{D}_s(t), \mathbf{D}_c(t)] = \left[\mathbf{d}_{1}^{s}(t),\dots,\mathbf{d}_{M}^{s}(t),\mathbf{d}_{1}^{c}(t),\dots,\mathbf{d}_{N}^{c}(t)\right]$ with the constraint
\begin{equation}\label{Scen1,Constraint}
\sum_{m=1}^{M} d_{m, l}^{s}(t) + \sum_{n=1}^{N} d_{n, l}^{c}(t)\leq 1, \forall l \in \mathcal{L}.
\end{equation}
Each device uses full power for transmission. 
Given the transmission power budget $P_{\max}$ and the receiving noise power $\sigma^2$, the receiving SNR of sensor $m$ and actuator $n$ is
\begin{equation}\label{Scen1,SINR}
\gamma_{m}^{s}(t)\!=\!\frac{\mathbf{d}_{m}^{s}(t)^{\top}\!\mathbf{g}_{m}^{s}(t)P_{\max }}{\sigma^{2}}, \gamma_{n}^{c}(t)\!=\!\frac{\mathbf{d}_{n}^{c}(t)^{\top}\!\mathbf{g}_{n}^{c}(t)P_{\max }}{\sigma^{2}}.
\end{equation}
Taking \eqref{Scen1,SINR} into \eqref{BLER}, the decoding error probability of a packet of sensor $m$ and actuator $n$ can be expressed as
\begin{equation}\label{eq:PDR}
{\varepsilon}_m^{s}(t)\triangleq\varepsilon(\gamma_m^{s}(t)), {\varepsilon}_n^{c}(t)\triangleq\varepsilon(\gamma_n^{c}(t)).
\end{equation}
Then we have $\operatorname{Pr}[\phi_{m}(t) = 1] = 1 - {\varepsilon}_m^{s}(t)$ and $\operatorname{Pr}[\lambda_{n}(t) = 1] = 1 - {\varepsilon}_n^{c}(t)$. 

\subsection{State Estimation} \label{sec:Estimator}
Due to the presence of noise, represented by $\mathbf{v}(t)$ and $\mathbf{w}(t)$, and the partial observability introduced by the matrix $\mathbf{C}$, along with imperfect communication (such as decoding errors and limited channel access) of sensing packets, the true plant state $\mathbf{x}(t)$ is not directly observable by the edge controller. Instead, the controller must rely on an estimated state, $\hat{\mathbf{x}}(t)$. Therefore, a state estimator is implemented on the controller side.

We assume that the edge controller has complete knowledge of the constant system matrices $\mathbf{A}, \mathbf{B}, \mathbf{C}, \mathbf{W}$, and $\mathbf{V}$, and can detect whether received sensing packets are successfully decoded through cyclic redundancy check (CRC), indicated by $\mathbf{\Psi}(t)$. Additionally, we assume that the controller receives perfect feedback from the actuators regarding the success of control packet transmissions, captured by $\mathbf{\Lambda}(t)$. This ensures that the controller knows both the received control inputs $\mathbf{u}^{\mathrm{rx}}(t)$ and sensor measurements $\mathbf{y}^{\mathrm{rx}}(t)$. Given this information, the controller performs optimal state estimation using a modified Kalman filter (mKF) with partially observed sensing measurements \cite{Peters2016Codesign}, formulated as follows:
\begin{subequations} \label{eq:KF}
\begin{align}
\hat{\mathbf{x}}_{\mathrm{est}}^{-}(t) \! \!&\triangleq \!\!\mathbf{A} \hat{\mathbf{x}}_{\mathrm{est}}(t-1) + \mathbf{B} \mathbf{u}^{\mathrm{rx}}(t-1)\!,  \label{KF,a}\\
\mathbf{P}_{\mathrm{est}}^{-}(t)\!\! &=\!\!\mathbf{A} \mathbf{P}_{\mathrm{est}}(t-1) \mathbf{A}^\top+\mathbf{W}\!,\label{KF,b}\\
\mathbf{K}_{\mathrm{est}}(t) \!\!&=\!\!\mathbf{P}_{\mathrm{est}}^{-}(t)\mathbf{C}^\top\!\mathbf{\Psi}(t)\!^\top\!\!\left(\!\mathbf{\Psi}(t)\mathbf{C}\mathbf{P}_{\mathrm{est}}^{-}(t)\mathbf{C}^\top\mathbf{\Psi}(t)\!^\top\!\!\!+\!\!\mathbf{V}\!\right)\!\!^{-\!\mathbf{1}}\!,\label{KF,c}\\
\hat{\mathbf{x}}_{\mathrm{est}}(t) \! \!&\triangleq\! \!\hat{\mathbf{x}}_{\mathrm{est}}^{-}(t)+\mathbf{K}_{\mathrm{est}}(t)\left(\mathbf{y}^{\mathrm{rx}}(t)-\mathbf{\Psi}(t)\mathbf{C}\hat{\mathbf{x}}_{\mathrm{est}}^{-}(t)\right)\!,\label{KF,d}\\
\mathbf{P}_{\mathrm{est}}(t) \!\!&=\!\!\left(\mathbf{I}-\mathbf{K}_{\mathrm{est}}(t) \mathbf{\Psi}(t) \mathbf{C}\right) \mathbf{P}_{\mathrm{est}}^{-}(t)\!,\label{KF,e}
\end{align}
\end{subequations}
where $\mathbf{I}$ is the $M\times M$ identity matrix; $\hat{\mathbf{x}}_{\mathrm{est}}^{-}(t)$ is the prior state estimate; $\hat{\mathbf{x}}_{\mathrm{est}}(t)$ is the posterior state estimate of $\mathbf{x}(t)$ at time $t$, i.e., the pre-filtered measurement of $\mathbf{y}^{\mathrm{rx}}(t)$; $\mathbf{K}_{\mathrm{est}}(t)$ is the time-varying Kalman gain; The matrices $\mathbf{P}_{\mathrm{est}}^{-}(t)$ and $\mathbf{P}_{\mathrm{est}}(t)$ represent the prior and posterior error covariance at the sensor at time $t$, respectively. \eqref{KF,a} and \eqref{KF,b} present the prediction steps while \eqref{KF,c}, \eqref{KF,d}, and \eqref{KF,e} correspond to the updating steps.

At the beginning of each time slot, the controller generates control commands based on the current plant state estimation and transmits them within the same time slot. However, due to a one-step transmission delay, the sensing packets measuring the current plant state are only received at the end of the time slot. As a result, the controller must rely on the state estimation from the mKF obtained in the previous time slot. Therefore, to estimate the current plant state, the controller updates the current estimation by
\begin{equation}\label{eq:Estimation}
\hat{\mathbf{x}}(t) \!\triangleq\! \mathbf{A}\hat{\mathbf{x}}_{\mathrm{est}}(t\!-\!1)\!+\!\mathbf{B}\mathbf{u}^{\mathrm{rx}}(t\!-\!1).
\end{equation}
The corresponding estimation error covariance matrix $\mathbf{P}(t)$ is
\begin{equation}\label{eq:EstimationP}
\begin{aligned}
    \mathbf{P}(t) \!&\triangleq \mathbb{E}\left[\left(\hat{\mathbf{x}}(t)-\mathbf{x}(t)\right)\left(\hat{\mathbf{x}}(t)-\mathbf{x}(t)\right)^\top\right], \\
    &= \! \mathbf{A}\mathbf{P}_{\mathrm{est}}(t\!-\!1)\!\mathbf{A}\!^\top\!\! +\! \mathbf{W}.
\end{aligned}
\end{equation}

\subsection{Performance Metrics}
We focus on the discounted control cost of \eqref{LTI} defined as
\begin{equation} \label{eq:Long-termCost}
J\!=\! \lim _{T \rightarrow \infty}\mathbb{E}\!\!\left[\sum_{t=0}^{T-1}\!\beta^t\!\!\left(\mathbf{x}(t)\!^\top\mathbf{Q}\mathbf{x}(t)\!+\!\mathbf{u}^{\mathrm{rx}}(t)\!^\top\mathbf{R}\mathbf{u}^{\mathrm{rx}}(t)\right)\right],
\end{equation}
where $\beta \in (0,1]$ is a discount factor, and $\mathbf{Q}$ and $\mathbf{R}$ are positive semi-definite weighting matrices for the state and control input, respectively. The discount factor introduces a time-weighting mechanism, allowing flexibility in controlling the trade-off between short-term responsiveness and long-term performance. The control cost in \eqref{eq:Long-termCost} reflects the deviation of the plant state $\mathbf{x}(t)$ from desired states (assumed to be zero for simplicity) and the energy required for control inputs $\mathbf{u}^{\mathrm{rx}}(t)$. Our goal is to minimize these deviations and do so efficiently by reducing the overall control cost.

By leveraging the state estimation $\hat{\mathbf{x}}(t)$ in \eqref{eq:Estimation} and the estimation error covariance matrix $\mathbf{P}(t)$ in \eqref{eq:EstimationP}, we have the following equality
\begin{equation} \label{eq:RewardLemma}
        \mathbb{E}[\mathbf{x}(t)\!^\top\!\mathbf{Q}\mathbf{x}(t)] = \hat{\mathbf{x}}(t) ^\top\!\mathbf{Q}\hat{\mathbf{x}}(t) + \operatorname{Tr}(\mathbf{Q}\mathbf{P}(t)).
\end{equation}
By substituting \eqref{eq:RewardLemma} into \eqref{eq:Long-termCost}, it becomes clear that minimizing the control cost inherently requires reducing the estimation error $\mathbf{P}(t)$. Thus, the objective of minimizing the control cost translates into finding an optimal codesign policy that keeps the system close to its desired state while efficiently managing the estimation error caused by noise and uncertainties in wireless communications. This control cost framework aligns with the goals of the codesign problem because it addresses two critical factors: 1) balancing communication resources between sensing and control to maintain system observability and control accuracy, and 2) generating intelligent control inputs that efficiently regulate the system state.

\subsection{Codesign Policies}\label{sec:CodesignPolicies}
The edge controller implements a codesign policy that dynamically schedules system devices and generates control commands based on the available information from the WNCS. This codesign policy maps the system state (i.e., the available information) to an action, which can be expressed as:
\begin{equation} \label{eq:CoDesignPolicy}
    (\mathbf{D}(t),\mathbf{u}^{\mathrm{tx}}(t)) = \pi(\mathbf{s}(t)),
\end{equation}
where $\pi(\cdot)$ is the codesign policy that generates the overall channel allocation $\mathbf{D}(t)$ for all devices and the control inputs $\mathbf{u}^{\mathrm{tx}}(t)$ for all actuators; $\mathbf{s}(t)$ is the system state and will be discussed in the next section. 

In the WNCS, uplink (sensor measurements) and downlink (control signals) transmissions share limited wireless resources. The control cost depends on both the accuracy of state estimation (influenced by uplink transmissions) and the effectiveness of control actions (dependent on downlink transmissions). As a result, decisions in communication and control are interdependent. For example, the control policy requires timely, accurate sensor data, while the communication policy must prioritize transmissions based on the system’s control demands. Optimizing communication without considering control requirements can lead to inefficiencies, such as over-allocating resources to sensor updates when control actions are more critical.
Unlike traditional approaches that treat communication and control separately, the codesign policy in \eqref{eq:CoDesignPolicy} jointly optimizes both, acknowledging their interdependencies. It ensures communication scheduling supports control needs, and control decisions are made inherently with communication limitations. In the next section, we formally define the codesign problem for the WNCS.

\section{Codesign Problem Formulation} \label{sec:Formulation}
We aim to find a deterministic and stationary codesign policy in \eqref{eq:CoDesignPolicy} that generates actions at each time slot to minimize the discounted control cost \eqref{eq:Long-termCost}. This involves jointly optimizing the objectives across estimation, control, and communication domains. The problem can be formulated as an MDP, as it is inherently a sequential decision-making task. The Markovian property holds clearly due to the Markov channel modeling in \eqref{ChannelTransition} and \eqref{eq:PDR} as well as the updating rule in \eqref{LTI} and \eqref{eq:Estimation}. In the sequel, we present the MDP formulation for the codesign problem.

\subsection{MDP Formulation}
\subsubsection{States} 
Given the estimated plant state $\hat{\mathbf{x}}(t) \in \mathbb{R}^{K}$, the estimation quality state $\mathbf{P}(t) \in \mathbb{R}^{K\times K}$, the control quality state $\mathbf{u}^{\mathrm{rx}}(t-1) \in \mathbb{R}^{N}$, and the CSI state $\mathbf{G}(t) \in \mathcal{G}^{L \times (M+N)}$, the MDP state set is defined as $\mathbf{s}(t) \triangleq \{\hat{\mathbf{x}}(t),\mathbf{P}(t),\mathbf{u}^{\mathrm{rx}}(t-1),\mathbf{G}(t)\} \in \mathbb{R}^{K} \times \mathbb{R}^{K\times K} \times \mathbb{R}^{N} \times \mathcal{G}^{L \times (M+N)}$. Integrating the estimated plant state and the estimation quality state enables the policy to be aware of the sensing and control performance. The inclusion of a control quality state helps agents make smart control decisions by making full use of the effective control efforts in the last time slot. This can justify the effectiveness of the previously generated control inputs, evaluate their impacts on the control system, and estimate the control availability, making better-informed control decisions. The CSI state determines the channel quality and affects the updating of the estimation quality state, the control quality state, and the estimated plant state, thereby impacting the control cost.

\subsubsection{Actions} 
Given the generated control input $\mathbf{u}^{\mathrm{tx}}(t) \in \mathbb{R}^{N}$ and the channel allocation $\mathbf{D}(t) \in \{0,1\}^{(N+M) \times L}$  under constraint \eqref{Scen1,Constraint}, the MDP action set is defined as $\mathbf{a}(t) \triangleq \{\mathbf{D}(t),\mathbf{u}^{\mathrm{tx}}(t)\} \in \{0,1\}^{(N+M) \times L} \times \mathbb{R}^{N}$. In the codesign problem, the action allocates $L$ frequency channels to $(M+N)$ devices and decides the continuous control inputs for $N$ actuators. The discrete action space is defined by the number of all possible allocations, which is denoted as $\mathcal{A}_d$ with the cardinality of $\left|\mathcal{A}_d\right|=\sum_{l=1}^{L} \frac{L!}{l!(L-l)!} \frac{(M+N)!}{(M+N-l)!}$. The continuous action space for control is defined as $\mathcal{A}_c \triangleq \mathbb{R}^{N}$. Thus, the hybrid actions space is denoted as $\mathcal{A} \triangleq \mathcal{A}_d \times \mathcal{A}_c$.

\subsubsection{Transitions} 
Let $\mathbf{s}^{+} (t) \triangleq \left\{\hat{\mathbf{x}}(t),\mathbf{P}(t),\mathbf{u}^{\mathrm{rx}}(t-1)\right\}$ represent the state set excluding the CSI. $\mathbf{s}^{+} (t)$ and $\mathbf{G}(t)$ are independent. Hence, the state-transition probability from $\mathbf{s}(t)$ to $\mathbf{s}(t+1)$ under a particular action $\mathbf{a}(t)$ is
\begin{equation} \label{eq:MDPtransition}
\begin{aligned}
    &\operatorname{Pr}[\mathbf{s}(t+1) \mid \mathbf{s}(t), \mathbf{a}(t)] \\
    &=\operatorname{Pr}[\mathbf{G}(t+1) \mid \mathbf{G}(t)] \operatorname{Pr}[\mathbf{s}^{+}(t+1) \mid \mathbf{s}(t), \mathbf{a}(t)],
\end{aligned}
\end{equation}
where $\operatorname{Pr}[\mathbf{G}(t\!+\!1)\!\!\mid\!\!\mathbf{G}(t)]$ is the transition probability of the channel state and can be directly obtained from the channel state transmission matrices $[\mathbf{M}]_{i,j}$ in \eqref{ChannelTransition}. $\operatorname{Pr}[\mathbf{s}^{+}(t\!+\!1)\!\!\mid\!\!\mathbf{s}(t),\mathbf{a}(t)]$ in \eqref{eq:MDPtransition} is related to the transition probability of the estimated plant state, the estimation quality state, and the control quality state, which can be obtained directly from the decoding failure probability in \eqref{eq:PDR} and the updating rule in \eqref{LTI} and \eqref{eq:Estimation}.

\subsubsection{Costs} 
By substituting \eqref{eq:RewardLemma} into \eqref{eq:Long-termCost}, \eqref{eq:Long-termCost} can be rewritten as
\begin{equation}\label{eq:MDPproblem}
J^*\!\! = \!\!\min_{\pi(\cdot)}\!\lim _{T\rightarrow \infty}\!\mathbb{E}\!\!\left[\sum_{t=0}^{T-1}\!\beta^tc(t)\right]\!\!,
\end{equation}
where $c(t) = \!\!\hat{\mathbf{x}}(t)\!^\top\!\mathbf{Q}\hat{\mathbf{x}}(t)\!+\!\operatorname{Tr}(\mathbf{Q}\mathbf{P}(t))\!+\!\mathbf{u}^{\mathrm{rx}}(t)\!^\top\!\mathbf{R}\mathbf{u}^{\mathrm{rx}}(t)$ is one-step cost. 
The one-step cost $c(t)$ penalizes the error of the estimated plant state, the state estimation error, and control efforts at each time slot, respectively. The MDP solution aims to find a codesign policy for minimizing the total cost $J^*$ in \eqref{eq:MDPproblem} accumulating one-step costs $c(t)$ over time with a discount factor $\beta$. The edge controller fully observes the MDP state by the assumptions presented in Section~\ref{sec:sys} and can calculate the one-step cost at each time slot.

\subsection{Challenges for DRL-based Codesign} \label{sec:ChallengesDRL}
The formulated MDP problem motivates the utilization of DRL. On one hand, closed-form expressions between the cost function \eqref{eq:MDPproblem} and system state $\mathbf{s}(t)$ are analytically intractable. On the other hand, the formulated MDP problem is mixed-integer, nonlinear programming, and NP-hard, making it impossible to solve efficiently with traditional methods for solving MDPs, such as value iteration, policy iteration, and linear programming. These challenges necessitate the development of advanced DRL approaches to address the codesign MDP problem. However, the following characteristics of the codesign MDP pose challenges to DRL-based solutions.

\subsubsection{Large and hybrid action space} The formulated MDP presents a large and hybrid action space $\mathcal{A}$, which poses a significant challenge for conventional single-agent DRL approaches, typically designed to handle either discrete or continuous action spaces, but not both simultaneously. A common approach is to decouple the hybrid action space using two separate DRL agents, one responsible for discrete actions and the other for continuous actions, forming a hierarchical or dual-agent DRL framework \cite{Zhao2023IoTJ,funk2021learning,kesper2023toward}. However, this introduces a non-stationary environment, making learning unstable and convergence difficult, as the interaction between agents destabilizes the training process \cite{Termehchi}. Moreover, this approach suffers from the curse of dimensionality as the system scales, especially in the discrete action space, even with modest numbers of devices and channels. For example, when $M=N=L=5$, the MDP has 63,590 discrete actions for channel assignment and 5 continuous actions for control. For $M=N=L=10$, the number of discrete actions increases to approximately $15 \times 10^{11}$, with 10 continuous control actions.

This explosion in action space makes training a DRL agent extremely challenging, requiring DNNs with large storage and high computational complexity. As a result, current communication-control codesign methods are limited to small system scales. For instance, prior works \cite{Zhao2023IoTJ,funk2021learning,kesper2023toward} only consider two discrete actions, i.e., whether or not to communicate. One potential solution is to approximate the large discrete action space with continuous values, allowing a single DRL agent to handle both discrete and continuous actions. This is feasible for actions with some continuity, such as discrete time intervals \cite{wan2023model,wan2023integrated,wang2023deep}, but not for channel assignment, where ``continuity" lacks physical meaning. Alternative approaches, such as threshold triggers \cite{9561274,shibata2023deep} or priority mechanisms \cite{8619335,PangFBL}, introduce a mapping between continuous outputs and discrete actions, but they often lead to unstable training and suboptimal policies due to the non-smooth nature of the mapping.

\subsubsection{Correlated communication and control policies}
In DRL-based codesign, a significant challenge lies in effectively capturing the correlation between the communication policy and the control policy. Although the one-step cost function in the codesign problem accounts for the joint effects of both policies, a basic DRL algorithm may struggle to learn the complex interactions between them. This is because the cost function reflects the outcome of the chosen actions, but it does not explicitly model the underlying dependencies between communication and control. As a result, the DRL agent might only indirectly learn the relationship between the two policies through the scalar cost, which can lead to suboptimal policy learning. Without a mechanism to directly integrate these interdependencies, the DRL agent may fail to fully optimize the system performance.

One approach to address the correlation between communication and control policies is to use two separate actors or dual agents, with each agent specializing in one policy \cite{Zhao2023IoTJ,funk2021learning,kesper2023toward}. However, this design lacks a direct mechanism for sharing information between scheduling and control decisions, limiting the ability to fully capture their interdependencies. Alternatively, a single-branch actor that encodes both policies with a shared representation \cite{wan2023model,wan2023integrated,wang2023deep,9561274,shibata2023deep,8619335} improves coordination but weakens the ability to handle the specific needs of each policy domain—communication and control. The main challenge is to create a mechanism that allows efficient information sharing between the two policies while ensuring that their unique roles—managing communication and control—are effectively optimized together.

\subsubsection{Cross-domain variables and floating control inputs}
In the MDP state space, we encounter variables from different domains: the estimated plant state $\hat{\mathbf{x}}(t)$ from control, the estimation quality state $\mathbf{P}(t)$ from estimation, and the channel quality state $\mathbf{G}(t)$ from communication. Similarly, the action space includes continuous control inputs and discrete channel allocation decisions, spanning these different domains. This heterogeneity in cross-domain input and output variables presents significant challenges for DRL due to their differing scales, units, and coupled dynamics.

DNNs, which are employed in DRL, assume input variables are on similar scales and distributions. The varying scales of cross-domain variables can disrupt training, as certain variables may dominate the learning process, leading to skewed gradient updates during backpropagation. This imbalance hinders the DNN’s ability to learn an effective policy across all domains. Additionally, the interdependencies between control, estimation, and communication mean that a change in one domain can propagate and affect others, further complicating the learning process. For large-scale WNCSs with numerous variables, it is crucial that the DRL model accurately captures these interactions while maintaining balanced sensitivity across all domains in both the state and action spaces.

Another issue arises with floating control inputs, where the DRL may output control signals for actuators that have not been scheduled during training. These floating actions are ineffective because they do not influence the system's behavior, as the corresponding actuators are inactive. However, they still introduce noise and irrelevant data into the training process. The presence of these unused actions can mislead the learning algorithm by adding unnecessary complexity, making it harder for the DRL model to identify meaningful patterns between actions and their actual effects on the system. This noise can distort the gradients during backpropagation, slowing down convergence and reducing the effectiveness of policy updates. In particular, it complicates the learning process by increasing the variance in the training data, making it more challenging to learn a robust and efficient policy. These challenges have not been adequately addressed in existing works \cite{8619335,funk2021learning,kesper2023toward,9561274,shibata2023deep,wan2023model,wan2023integrated,wang2023deep,redder2019deep,Termehchi,Zhao2023IoTJ}.

\section{DRL Algorithm for the Codesign Problem} \label{sec:DRL}
\begin{figure}[t]
    \centering
    \includegraphics[width=3.5in]{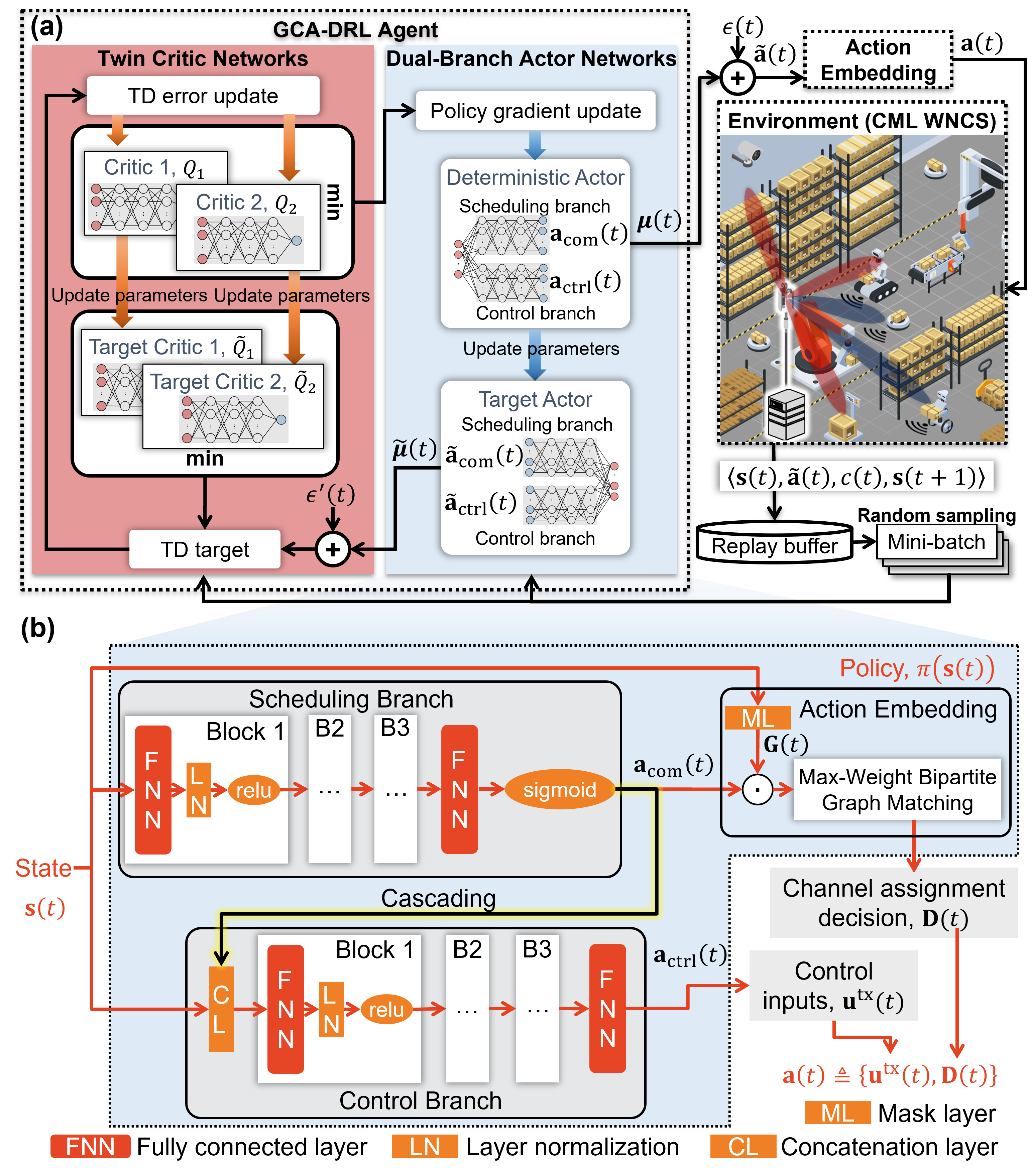}
    \vspace{-0.8cm}
    \caption{Illustration of GCA-DRL: (a) Learning structure. TD denotes temporal difference. (b) Dual-branch actor and action embedding scheme.} 
    \label{fig:ADAGE-DRL}
    \vspace{0.0cm}
\end{figure}

To address the challenges outlined in Section~\ref{sec:ChallengesDRL}, we introduce the novel GCA-DRL algorithm, as illustrated in Fig.~\ref{fig:ADAGE-DRL}. GCA-DRL is a deterministic DRL algorithm that builds upon the Twin Delayed Deep Deterministic Policy Gradient (TD3) algorithm \cite{fujimoto2018addressing}. TD3 is an actor-critic method known for mitigating overestimation bias and improving learning stability—two key issues in dynamic, high-variance environments. Compared to other DRL methods, TD3's twin critic architecture and delayed policy updates make it more robust against the overestimation and instability that can arise in complex decision-making problems, including the codesign problem. These advantages enable TD3 to handle intricate interdependencies between communication and control policies more effectively than alternatives, such as Deep Deterministic Policy Gradient (DDPG) or Proximal Policy Optimization (PPO), which may struggle with biased learning and unstable convergence in such environments. In this section, we first present the fundamentals of the vanilla TD3 algorithm, followed by our specific innovations tailored to address the unique challenges of the codesign problem.

\subsection{Vanilla TD3 Algorithm}
\subsubsection{Basic Architecture}
The TD3 algorithm consists of two actor networks and two critic networks. The main actor network, $\pi_a(\mathbf{s}(t); \theta)$, maps the current state $\mathbf{s}(t)$ to an action $\mathbf{\mu}(t)$, where $\theta$ represents the network parameters. To encourage exploration during training, Gaussian noise $\epsilon(t)\!\sim\!\mathcal{N}(0, \sigma_a^2)$ is added to the action, resulting in $\tilde{\mathbf{a}}(t)\!=\!\mathbf{\mu}(t)\!+\!\epsilon(t)$. The objective is to minimize the long-term cumulative cost, defined in \eqref{eq:MDPproblem}, by selecting actions that reduce the overall cost over time.

The two critic networks, $Q_1(\mathbf{s}(t), \tilde{\mathbf{a}}(t); \varphi_1)$ and $Q_2(\mathbf{s}(t), \tilde{\mathbf{a}}(t); \varphi_2)$, estimate the expected cumulative cost for each state-action pair and provide feedback to the main actor. These networks use different sets of parameters, $\varphi_1$ and $\varphi_2$, to prevent overestimation bias—a common problem in single-critic architectures, where the critic tends to overestimate the value of certain actions. By using two critics, TD3 mitigates this bias by taking the minimum of the two Q-values, ensuring more conservative and stable updates. This helps guide the actor to select actions that consistently minimize long-term costs.  The Q-value for each critic is:
\begin{equation}
Q_k(\mathbf{s}(t),\tilde{\mathbf{a}}(t);\varphi_k)\!\approx\mathbb{E}\!\left[\sum_{i=0}^{\infty} -\beta^{i} c(t\!+\!i) \bigg|\mathbf{s}(t),\tilde{\mathbf{a}}(t),\varphi_k\right]\!\!.
\end{equation}

TD3 also employs a target actor network, $\hat{\pi}(\mathbf{s}(t); \hat{\theta})$, and two target critic networks, $\hat{Q}_1$ and $\hat{Q}_2$, to improve stability during training. The target networks are slow-moving copies of the main networks, and they help stabilize learning by providing more consistent target values during updates, reducing the likelihood of sudden shifts in policy. The target actor generates actions $\hat{\mathbf{a}}(t)$ by adding clipped noise, and the critics compute Q-values. Critic updates are performed using the minimum of the two Q-values from the target critics, which helps reduce overestimation bias and ensures conservative updates.

Both the actor and critic networks are fully connected, feedforward structures with multiple hidden layers and ReLU activations. The main and target networks share similar architectures but are initialized with different parameters to reduce bias and ensure robustness during learning.

\subsubsection{Training Algorithm}

The training process of TD3 alternates between experience generation and policy updates. Below, we outline the key steps and high-level concepts of TD3.
Using the actor network $\pi_a(\cdot; \theta)$, the TD3 agent continuously interacts with the environment, sampling transitions in the form of $(\mathbf{s}(t), \tilde{\mathbf{a}}(t), c(t), \mathbf{s}(t+1))$, which are stored in a replay buffer. The TD3 agent then randomly samples a mini-batch of $B$ transitions from the replay buffer, represented as $<\mathbf{s}_i, \tilde{\mathbf{a}}_i, c_i, \mathbf{s}_i^{'}>$, for each $i \in \{1, 2, \dots, B\}$.
We define the temporal-difference (TD) error for each transition as:
\begin{equation}\label{eq:TDerror}
    \mathrm{TD}_i^{k} \triangleq y_i - Q_k(\mathbf{s}_i, \tilde{\mathbf{a}}_i; \varphi_k), \forall k \in \{1,2\},
\end{equation}
where the TD target $y_i$ is
\begin{equation}\label{eq:TargetValue}
    y_i \triangleq -c_i + \beta \min_{k\in \{1,2\}} \hat{Q}_k(\mathbf{s}'_i,\hat{\pi}_a(\mathbf{s}'_i;\theta_t)+\epsilon';\hat{\varphi}_k).
\end{equation}

The TD3 agent updates the parameters of the critic networks by minimizing the following loss function across all sampled transitions:
\begin{equation} \label{eq:LossForCritic}
    L_k(\varphi_k) = \frac{1}{B} \sum_{i=1}^{B}{\mathrm{TD}_i^{k}}^2, \forall k \in \{1,2\},
\end{equation}
where $\mathrm{TD}_i^{k}$ is defined in \eqref{eq:TDerror}. 

Next, the actor network is updated by using the following policy gradient to minimize the expected cumulative cost:
\begin{equation} \label{eq:GradientForActorTD3}
\begin{aligned}
    \nabla_\theta L_a (\theta)&= \frac{1}{B}\sum_{i=1}^{B}G_{i}^aG_{i}^{\pi}, \\
    G_{i}^a &\triangleq \nabla_{\tilde{\mathbf{a}}_i}\min_{k\in \{1,2\}} Q_k(\mathbf{s}_i, \tilde{\mathbf{a}}_i; \varphi_k), \\
    G_{i}^{\pi} &\triangleq\nabla_\theta\pi_a(\mathbf{s}_i;\theta),
\end{aligned}
\end{equation}
where $G_i^a$ is the gradient of the minimum critic output with respect to the action $\tilde{\mathbf{a}}_i$, and $G_i^{\pi}$ is the gradient of the actor’s output with respect to the actor parameters. Both gradients are evaluated for each state $\mathbf{s}_i$.

Finally, the TD3 agent periodically updates the target networks’ parameters, $\hat{\theta}$ and $\hat{\varphi}_k$, with the latest parameters $\theta$ and $\varphi_k$ from the main networks, respectively.

\subsection{Proposed GCA-DRL}

\subsubsection{Cascaded dual-branch actor network for joint communication and control optimization} \label{sec:ActorDesign}
To address the challenges of correlated communication and control policies and the cross-domain variables in the codesign problem, we propose the GCA-DRL, designed with a dual-branch actor network that simultaneously tackles scheduling (communication) and control. The architecture outputs two types of decisions, $\mathbf{\mu}(t) = \left(\mathbf{a}_{\mathrm{com}}(t), \mathbf{a}_{\mathrm{ctrl}}(t)\right)$, where $\mathbf{a}_{\mathrm{com}}(t)$ represents the channel allocation (scheduling) and $\mathbf{a}_{\mathrm{ctrl}}(t)$ provides the control signals for the actuators. This structure allows for joint optimization by sharing critical information between the scheduling and control branches.

The scheduling branch outputs channel allocation decisions using a sigmoid activation function to produce $\mathbf{a}_{\mathrm{com}}(t) \in [0,1]^{(M+N) \times L}$, which ensures that the channel assignment is continuous, facilitating smooth learning. The control branch outputs continuous control signals $\mathbf{a}_{\mathrm{ctrl}}(t) \in \mathbb{R}^{N}$.

The novelty lies in the cascaded connection between the two branches: the scheduling output is passed into the control branch to inform control decisions based on the real-time availability of communication resources. This ensures that the control actions are fully aware of communication constraints, leading to joint optimization of communication and control policies. Without this cascading, control commands may fail to consider the actual communication state, leading to inefficiencies or even failed actions. By feeding the scheduling decisions into the control branch, GCA-DRL directly addresses the challenge of correlated communication and control policies, allowing both policies to align dynamically and enhance overall system performance.

Furthermore, the actor network adopts layer normalization (LN) across all fully connected layers to handle the challenge of cross-domain variables. Cross-domain inputs, including communication quality, control states, and estimation states, often have differing scales. Let's assume the output of the fully connected layer is denoted by a vector $\mathbf{z} = (z_1,\dots,z_d)\in \mathbb{R}^{d}$. We have the mean $\mu_z = \frac{1}{d}\sum_{i=1}^{d}z_i$ and variance $\sigma_z^2 = \frac{1}{d}\sum_{i=1}^{d}(z_i-\mu_z)^2$ of the elements in $\mathbf{z}$. Each element of $\mathbf{z}$ is then normalized by 
\begin{equation}\label{eq:LN}
    \hat{z}_i = \alpha_i \frac{z_i - \mu_z}{\sqrt{\sigma_z^2 + \epsilon_z}}+\delta_i, \forall i\in\{1,2,\dots,d\}
\end{equation}
where $\alpha_i$ is a learnable scale parameter; $\delta_i$ is a learnable shift parameter; $\epsilon_z$ is a small constant added for numerical stability. LN ensures that each input to the network is normalized, preventing certain variables from dominating the learning process and enabling more effective training across domains. This is essential for large-scale WNCSs, where maintaining the balance between communication and control domains is crucial for learning an effective policy. 

\subsubsection{Knowledge-assisted graph-based action embedding for effective communications}\label{sec:GAE}
To address the challenge of large hybrid action spaces, GCA-DRL leverages max-weight bipartite graph matching (MBGM) theory to transform the discrete channel allocation decisions into a continuous representation \cite{duff2001algorithms}. This helps reduce the complexity of the action space in DRL training, enabling more efficient learning.

In MBGM, we model the devices (sensors and actuators) and communication channels as two distinct sets of nodes in a bipartite graph. The edges between these nodes are assigned weights that represent the utility of assigning a particular device to a particular channel. The goal of MBGM is to find a one-to-one matching between devices and channels that maximizes the total utility, which is calculated based on these edge weights. This matching problem inherently incorporates the constraint \eqref{Scen1,Constraint} and can be solved using a well-established algorithm as described in \cite{duff2001algorithms}.

To map the discrete channel allocation decisions $\mathbf{D}(t)$ into continuous values, the actor network generates a weight matrix $\mathbf{a}_{\mathrm{com}}(t)$ through the scheduling branch, which is passed through a sigmoid activation function to ensure the values lie between 0 and 1. These weights represent the likelihood or importance of assigning a device to a channel.

By representing the channel allocation decisions as continuous values, we reduce the burden of exploring a large discrete action space during DRL training. Instead of the DRL agent needing to evaluate a vast number of discrete possibilities, it can explore continuous weight values incrementally, allowing for smoother exploration and faster convergence. This method avoids abrupt shifts in decision-making, which are common when directly handling discrete actions, and instead enables incremental adjustments during training.

A key innovation in GCA-DRL is the incorporation of domain-specific knowledge into the MBGM process. Specifically, the actor network adjusts the weights in MBGM using CSI to make the scheduling decisions adaptive to real-time communication conditions. This is achieved by taking the element-wise product of the output from the scheduling branch and the CSI, i.e., $\mathbf{a}_{\mathrm{com}}(t) \odot \mathbf{G}(t)$. By multiplying the scheduling weights with CSI, the resulting weights become context-aware, ensuring that devices are assigned to channels with favorable conditions.
For instance, if a channel has poor quality (low CSI), its weight is reduced, lowering the chance of selection. This helps the DRL agent prioritize better channels, improving communication reliability. By focusing on channels with good conditions, the action space is pruned, avoiding poor performance options.

By using real-time CSI in the MBGM process, GCA-DRL improves training efficiency and decision quality. It reduces the exploration of irrelevant options, allowing the agent to focus on more promising actions, making learning faster and more effective by exploiting known CSI.

\subsubsection{Training with partial policy gradients update for effective control}\label{sec:Training}
To address the challenge of floating control inputs—control signals generated for unscheduled actuators that introduce noise during training—GCA-DRL employs a partial policy gradient update mechanism. This mechanism focuses solely on the effective control inputs, i.e., those associated with scheduled actuators, thereby eliminating irrelevant signals and improving the stability of the training process.

The effective control inputs are defined as $\mathbf{a}_{\mathrm{ctrl}}^{i,\mathrm{eff}}$, which filters the raw control inputs $\mathbf{a}_{\mathrm{ctrl}}^i$ through the actuator scheduling matrix $\mathbf{D}_c^i$. This filtering removes control inputs for unscheduled actuators, reducing irrelevant data that can disrupt learning. The valid action set for the actor parameters update is $\tilde{\mathbf{a}}_v^i = (\mathbf{a}_{\mathrm{com}}^i, \mathbf{a}_{\mathrm{ctrl}}^{i,\mathrm{eff}})$.

The actor parameters are then updated using the partial gradient of the minimum critic output with respect to the effective actions, ensuring that only valid actions are used in the optimization process: 
\begin{equation} \label{eq:GradientForActor}
\nabla_\theta L_a (\theta)= \frac{1}{B}\sum_{i=1}^{B}\nabla_{\tilde{\mathbf{a}}_v^i}\min_{k\in \{1,2\}} Q_k(\mathbf{s}_i, \tilde{\mathbf{a}}_i; \varphi_k)G_{i}^{\pi}. 
\end{equation} 
By focusing on effective control inputs, GCA-DRL eliminates the disruptive noise caused by floating control signals, leading to more stable training and faster convergence. It ensures that only valid control actions contribute to the policy updates. The training algorithm of GCA-DRL is shown in Algorithm~\ref{alg:training}.

\begin{algorithm}[t]
\small  
\caption{\small{GCA-DRL for communications-control codesign of a CML WNCS}}\label{alg:training}
    \begin{algorithmic}[1]
    \Require Episode number $E$, maximum steps per episode $T$, discount factor $\beta$, target smoothing factor $\tau$, mini-batch size $B$, the policy update frequency $D_1$, the target update frequency $D_2$, the standard deviation of Gaussian noise, $\sigma_a$.
    \Ensure Well-trained actor $\pi_a^*(\cdot)$.
    \State Initialize the actor $\pi_a(\cdot)$ with random parameter $\theta$ and initialize the target actor with the same random parameter $\hat{\theta}=\theta$.
    Initialize each critic $Q_k(\cdot)$ and each target critic $\hat{Q}_k(\cdot)$ with random parameter $\varphi_k$ and $\hat{\varphi}_k$, respectively. 
    \For{episode = 1,…,$E$}
        \State Initialize the plant state $\mathbf{x}(0) = \mathbf{0}$, the received control input $\mathbf{u}^{\mathrm{tx}}(0) = \mathbf{0}$, and randomly initialize the CSI, $\mathbf{G}(0)$.
        \For{$t$ = 0,…,$T$}
        \State For the current observation $\mathbf{s}(t)$, select action $\tilde{\mathbf{a}}(t) = \pi_a(\mathbf{s}(t);\theta) + \epsilon(t)$. Pass the action $\tilde{\mathbf{a}}(t)$ to the action embedding scheme to obtain the MDP action $\mathbf{a}(t)$.
        \State Execute the MDP action $\mathbf{a}(t)$ into the CML WNCS defined in \eqref{LTI} with the estimator defined in \eqref{eq:Estimation}. 
        \State Observe the cost $c(t)$ and the next observation $\mathbf{s}(t+1)$.
        \State Store the experience $<\mathbf{s}(t),\tilde{\mathbf{a}}(t),c(t),\mathbf{s}(t+1)>$ in the replay buffer.\label{alg:experiences}
        \State Sample a random mini-batch of $B$ experiences $<\mathbf{s}_i,\tilde{\mathbf{a}}_i,c_i,\mathbf{s}_i^{'}>, \forall i \in \{1,2,\dots,B\}$ from the replay buffer.
        \State Set the value function target $y_i$ according to \eqref{eq:TargetValue}.
        \State Update each critic parameter $\varphi_k$ by minimizing the loss function $L_k(\varphi_k)$ in \eqref{eq:LossForCritic}.
        \State Every $D_1$ steps, update the actor parameters $\theta$ by using the policy gradient $\nabla_\theta L_a (\theta)$ in \eqref{eq:GradientForActor}.
        \State Every $D_2$ steps, update the target parameters using the smoothing factor $\tau$, i.e., $\hat{\varphi}_k = \tau\varphi_k + (1-\tau)\hat{\varphi}_k$ and $\hat{\theta}= \tau\theta + (1-\tau)\hat{\theta}$.
        \EndFor
    \EndFor
    \end{algorithmic}
\end{algorithm}

\section{Numerical Simulation} \label{sec:simulation}
Our numerical experiments are implemented using  MATLAB 2024a based on the computing platform with Intel i7-13700k CPU @ 3.40 GHz and 32 GB RAM. GPU is not required. The details of the WNCS parameters and the GCA-DRL parameters are summarized in Table~\ref{tab:Setup}. The dynamic system matrices $\mathbf{A}$, $\mathbf{B}$, and $\mathbf{C}$ are randomly generated by leveraging the method presented in \cite{Leong2020OMA}.

\begin{table}[t]
\footnotesize
\setlength\tabcolsep{2pt}
\centering
\caption{Summary of Experimental Setup}
\vspace{-0.3cm}
\begin{tabular}{ll}
\hline\hline
\textbf{Items}   & \textbf{Value}    \\ \hline
\multicolumn{2}{l}{\textbf{The WNCS parameters}}  \\
\rowcolor[HTML]{EFEFEF} 
System matrices, $\mathbf{A}$, $\mathbf{B}$, $\mathbf{C}$  & Randomly generated \\
Eigenvalues of $\mathbf{A}$ matrix, $\rho_k(\mathbf{A})$ & $\rho_k(\mathbf{A}) \in (1.0, 1.1)$ \\
\rowcolor[HTML]{EFEFEF} 
Noise covariance matrices, $\mathbf{W}$, $\mathbf{V}$ & Identity matrix \\
Penalty matrices in control cost, $\mathbf{Q}$, $\mathbf{R}$ & Identity matrix \\
\rowcolor[HTML]{EFEFEF} 
Initial plant state, $\mathbf{x}(0)$ & $\mathbf{x}(0) = \mathbf{0}^\top$\\
Plant saturation state & $|\mathbf{x}_i(t)| \leq 100, i\in \mathcal{K}, \forall t>0$\\
\rowcolor[HTML]{EFEFEF} 
Transmit power budget {[}dBm{]}, $P_{\max}$    & 23    \\
Background noise power {[}dBm{]}, $\sigma^2$    & $-$60   \\
\rowcolor[HTML]{EFEFEF} 
Code rate {[}bps{]}, $b/l$    & 2        \\
Block length {[}symbols{]}, $l$    & 200     \\
\rowcolor[HTML]{EFEFEF} 
Channel states, $\mathcal{G}$  &  $\{10^{-10},10^{-9},\dots,10^{-1}\}$   \\
CSI transition matrix, $\{\mathbf{M}_{m,l}^{s}\}$, $\{\mathbf{M}_{n,l}^{c}\}$,  & Randomly generated \\

\hline

\multicolumn{2}{l}{\textbf{Training parameters of GCA-DRL}}  \\
\rowcolor[HTML]{EFEFEF} 
Episode number, $E$    &  3000    \\
Maximum time slots per episode, $T$     & 100      \\
\rowcolor[HTML]{EFEFEF} 
Hidden layer size & 300, 200, 100 \\
Learning rate of actor and critic    & 0.001    \\
\rowcolor[HTML]{EFEFEF} 
Optimizer during learning     & Adam      \\
Threshold value of the learning gradient & 1  \\ 
\rowcolor[HTML]{EFEFEF} 
Experience buffer length     & 100000     \\
Target smoothing factor, $\tau$   & 0.005    \\
\rowcolor[HTML]{EFEFEF} 
Mini-batch size, $B$    & 64     \\
Discount factor, $\beta$      & 0.99       \\
\rowcolor[HTML]{EFEFEF} 
Policy and target update frequency, $D_1$, $D_2$   & 2     \\
Standard deviation of actor action noise, $\sigma_a$  & $\sqrt{0.1}$   \\
\hline\hline
\end{tabular}
\label{tab:Setup}
\vspace{-0.1cm}
\end{table}

\subsection{Benchmark Policies}
\subsubsection{Control policies} 
We present two benchmark control algorithms of the linear quadratic regulator (LQR), which are widely adopted in the existing literature \cite{WNC1}.

\textbf{Standard LQR controller} is designed with perfect communication to minimize the control cost \eqref{eq:Long-termCost}. The generated control input at $t$ is given as
\begin{equation} \label{eq:StandardLQR}
\begin{aligned}
    \mathbf{u}^{\mathrm{tx}}(t) \!&\triangleq\! \mathbf{K}_{\mathrm{std}}\hat{\mathbf{x}}(t) \!=\! -\beta\!\left(\mathbf{B}^\top\mathbf{S}_\infty^{\mathrm{std}}\mathbf{B}\!+\!\mathbf{R}\right)^{\!-1\!}\mathbf{B}^\top\mathbf{S}_\infty^{\mathrm{std}}\mathbf{A}\hat{\mathbf{x}}(t),
\end{aligned}
\end{equation}
where the matrix $\mathbf{S}_\infty^{\mathrm{std}}$ can be computed recursively as follows:
\begin{equation}
\begin{aligned}
    \mathbf{S}_k^{\mathrm{std}} &\triangleq \beta\mathbf{A}^\top\mathbf{S}_{k\!+\!1}^{\mathrm{std}}\mathbf{A}\!+\!\mathbf{Q} \\
    &\!-\!\beta^2 \mathbf{A}^\top\mathbf{S}_{k\!+\!1}^{\mathrm{std}}\mathbf{B}\!\left(\mathbf{B}^\top\mathbf{S}_{k\!+\!1}^{\mathrm{std}}\mathbf{B}\!+\!\mathbf{R}\right)\!^{\!-\!1}\mathbf{B}^\top\mathbf{S}_{k\!+\!1}^{\mathrm{std}}\mathbf{A},
\end{aligned}
\end{equation}
with initial value $\mathbf{S}_0^{\mathrm{std}} = \mathbf{Q}$.

\textbf{Modified LQR controller} is designed with imperfect communication to minimize control cost \eqref{eq:Long-termCost}. Let $\mathbf{E} \in [0,1]^{N \times N}$ denote the diagonal matrix of control packet loss rates, where $[\mathbf{E}]_{n,n} \triangleq {\Bar{\varepsilon}}_n^{c}, \forall n \in \mathcal{N}$ is the average control packet loss rate of the actuator $n$. The generated control input at $t$ is given as
\begin{equation} \label{eq:ModifiedLQR}
\begin{aligned}
    \mathbf{u}^{\mathrm{tx}}(t) \!&\triangleq \!\mathbf{K}_{\mathrm{mod}}\hat{\mathbf{x}}(t)\! = \!-\beta\!\left(\mathbf{B}^\top\mathbf{S}_\infty^{\mathrm{mod}}\mathbf{B}\!+\!\mathbf{R}\right)^{\!-\!1\!}\mathbf{B}^\top\mathbf{S}_\infty^{\mathrm{mod}}\mathbf{A}\hat{\mathbf{x}}(t), \\
\end{aligned}
\end{equation}
where $\mathbf{S}_\infty^{\mathrm{mod}}$ can be computed recursively as follows:
\begin{equation}
\begin{aligned}
    \mathbf{S}_k^{\mathrm{mod}} &\triangleq \beta\mathbf{A}^\top\mathbf{S}_{k\!+\!1}^{\mathrm{mod}}\mathbf{A}\!+\!\mathbf{Q}\\
    &\!-\!\beta^2(\mathbf{I}\!-\!\mathbf{E}) \mathbf{A}^\top\mathbf{S}_{k\!+\!1}^{\mathrm{mod}}\mathbf{B}\!\left(\mathbf{B}^\top\mathbf{S}_{k\!+\!1}^{\mathrm{mod}}\mathbf{B}\!+\!\mathbf{R}\right)\!^{\!-\!1}\mathbf{B}^\top\mathbf{S}_{k\!+\!1}^{\mathrm{mod}}\mathbf{A},
\end{aligned}
\end{equation}
with initial value $\mathbf{S}_0^{\mathrm{mod}} = \mathbf{Q}$.
    
\subsubsection{Scheduling policies} \label{sec:BenchmarkSchedulers}
We present some benchmark scheduling algorithms that are commonly used in WNCSs.
\textbf{Round-robin policy} guarantees that no device is starved of channel access. All devices are numbered in a list. Channels are assigned to consecutive $L$ devices in the device list, shifting by one device each time slot. After reaching the last device, the allocation process wraps around to the start.
\textbf{Persistent policy} ensures that a device retains exclusive access to a frequency channel until its communication is successfully completed. Channels are initially assigned to $L$ devices that are randomly selected from all devices. Upon successful packet delivery of the scheduled device, the channel is then randomly reallocated to one of the unscheduled devices in the subsequent time slots.
\textbf{Greedy policy on age of information} (AoI) ensures timely data updates \cite{Leong2020OMA}. AoI of a device is defined as the time elapsed since the last successful reception of the packet by that device. At each time slot, devices select the remaining available channels based on their descend AoI ranking, choosing the channel with the best channel condition, until all channels are allocated.
\textbf{Greedy policy on CSI} uses the MBGM theory, where devices and channels are treated as two distinct sets of nodes. The edges between them have weights represented by the CSI between a particular device to a particular channel. The policy is to find a one-to-one matching that maximizes the total CSI at each time slot.
\textbf{Random policy} ensures that all devices have a statistically equal chance of accessing any of the channels over time. At each time slot, $L$ out of $M+N$ devices are randomly selected and each selected device is randomly allocated to one of the $L$ frequency channels.

We also compare our GCA-DRL with some codesign methods using DRL in the existing literature, which have been discussed in Section~\ref{sec:ChallengesDRL}.

\subsection{Performance Evaluation}
In Table~\ref{tab:PerformanceComparison}, we compare the control performance measured by the average sum of one-step costs, i.e., the control cost, among the proposed GCA-DRL algorithm and the benchmark approaches, when $K = M = N = L = 5$. There are 10 devices sharing 5 channels. Control costs in Table~\ref{tab:PerformanceComparison} are calculated by averaging 100 episodes with 100 steps per episode.
Notably, the developed GCA-DRL method demonstrates a significant advantage across all metrics compared to traditional designs and codesign benchmarks. Specifically, GCA-DRL achieves the lowest overall control cost of 858.65, which is a significant improvement over all other approaches. Compared to the best benchmark (1201.36), GCA-DRL reduces 28.5\% control cost. These results highlight the effectiveness of GCA-DRL in simultaneously optimizing communication and control policies, leading to better overall system performance as measured by the control cost.

\begin{table*}[t]
\centering
\footnotesize
\setlength\tabcolsep{3pt}
\caption{Comparison of the GCA-DRL and Benchmarks in the Control Cost $J$ with $K=M=N=L=5$.}
\vspace{-0.3cm}
\begin{tabular}{m{2.1cm}m{3.4cm}m{2.3cm}m{2.5cm}m{2.3cm}m{2.5cm}m{1.5cm}}
\hline\hline
\textbf{Design approach} &\textbf{Scheduler} & \textbf{Controller} & \textbf{State estimation cost, $\mathbb{E}[\operatorname{Tr}(\mathbf{Q}\mathbf{P}(t))]$} & \textbf{Control cost, $\mathbb{E}[\mathbf{u}^{\mathrm{rx}}(t)^\top\!\mathbf{R}\mathbf{u}^{\mathrm{rx}}(t)]$} & \textbf{Estimated state cost, $\mathbb{E}[\hat{\mathbf{x}}(t) ^\top\!\mathbf{Q}\hat{\mathbf{x}}(t)]$} & \textbf{Overall cost, $J$} \\ \hline\hline

\rowcolor[HTML]{EFEFEF} 
Traditional design &Random policy &  Modified LQR & 108.33 & 1897136202 & 6173988263 & 8071124573 \\ 
Traditional design &Random policy & Standard LQR & 108.33 & 8157.11 & 43777.50 & 52042.96  \\ 
\rowcolor[HTML]{EFEFEF} 
Traditional design &Random policy & Vanilla TD3 & 108.41 & 58.61  & 2337.41 &  2504.43  \\ 
Traditional design &Round-robin policy & Modified LQR & 134.69 & 7451458.74 & 32602309.72 & 40053903.15 \\ 
\rowcolor[HTML]{EFEFEF} 
Traditional design &Round-robin policy & Standard LQR & 134.69 & 16212.59 & 85268.12 & 101615.40  \\ 
Traditional design &Round-robin policy & Vanilla TD3 & 138.04 & 81.81  & 2902.85 &  3122.71 \\
\rowcolor[HTML]{EFEFEF} 
Traditional design &Persistent policy & Modified LQR & 99.41 & 16635697.50  & 43613727.11 &  60249524.02 \\ 
Traditional design &Persistent policy &  Standard LQR & 99.41 & 3671.22  & 21092.08 &  24862.72   \\ 
\rowcolor[HTML]{EFEFEF} 
Traditional design &Persistent policy & Vanilla TD3 & 98.18 & 64.78  & 1961.32 &  2124.28   \\ 
Traditional design &CSI-based greedy policy & Modified LQR & 61.57 & 15003542.91  & 34940948.69 &  49944553.17  \\ 
\rowcolor[HTML]{EFEFEF} 
Traditional design &CSI-based greedy policy & Standard LQR & 61.57 & 5723.14  & 27333.63 &  33118.35  \\ 
Traditional design &CSI-based greedy policy & Vanilla TD3 & 61.26 & 49.36  & 1570.45 &  1681.06 \\ 
\rowcolor[HTML]{EFEFEF} 
Traditional design &AoI-based greedy policy &  Modified LQR & 60.77 & 33229.38  & 145388.50 &  178678.67   \\ 
Traditional design &AoI-based greedy policy & Standard LQR & 60.77 & 442.66  & 2016.48 &  2519.92 \\ 
\rowcolor[HTML]{EFEFEF} 
Traditional design &AoI-based greedy policy & Vanilla TD3 & 60.75 & 80.18  & 1060.44 &  1201.36   \\ 
Traditional design &TD3 (priority-based mapping) & Standard LQR & 76.15 & 171.13  & 1554.48 &  1801.76  \\

\hline

\rowcolor[HTML]{EFEFEF} 
Codesign &TD3 (priority-based mapping) & Modified LQR & not converge & not converge & not converge  & not converge  \\ 
Codesign &TD3 (priority-based mapping) & Vanilla TD3 & 67.23 & 52.31 & 1071.68  & 1203.14 \\ 
\rowcolor[HTML]{EFEFEF} 
Codesign & \multicolumn{2}{l}{Single TD3 (rounded-off mapping and control)} & 222.08 & 130.36 & 5598.62  & 5598.62 \\ 
Codesign & \multicolumn{2}{l}{Single TD3 (threshold-based mapping and control)} & 116.73 & 58.77 &  1880.19 & 2055.68  \\ 
\rowcolor[HTML]{EFEFEF} 
Codesign & \multicolumn{2}{l}{Single TD3 (priority-based mapping and control)} & 66.99 & 58.02 & 1130.82  & 1255.82  \\ 
\textbf{Codesign} & \multicolumn{2}{l}{\textbf{GCA-DRL (This work)}} & \textbf{52.21} & \textbf{46.34} & \textbf{760.10} & \textbf{858.65} \\ \hline\hline
\end{tabular}
\label{tab:PerformanceComparison}
\vspace{-0.5cm}
\end{table*}

\begin{figure}[t]
	\centering\includegraphics[width=3.1in]{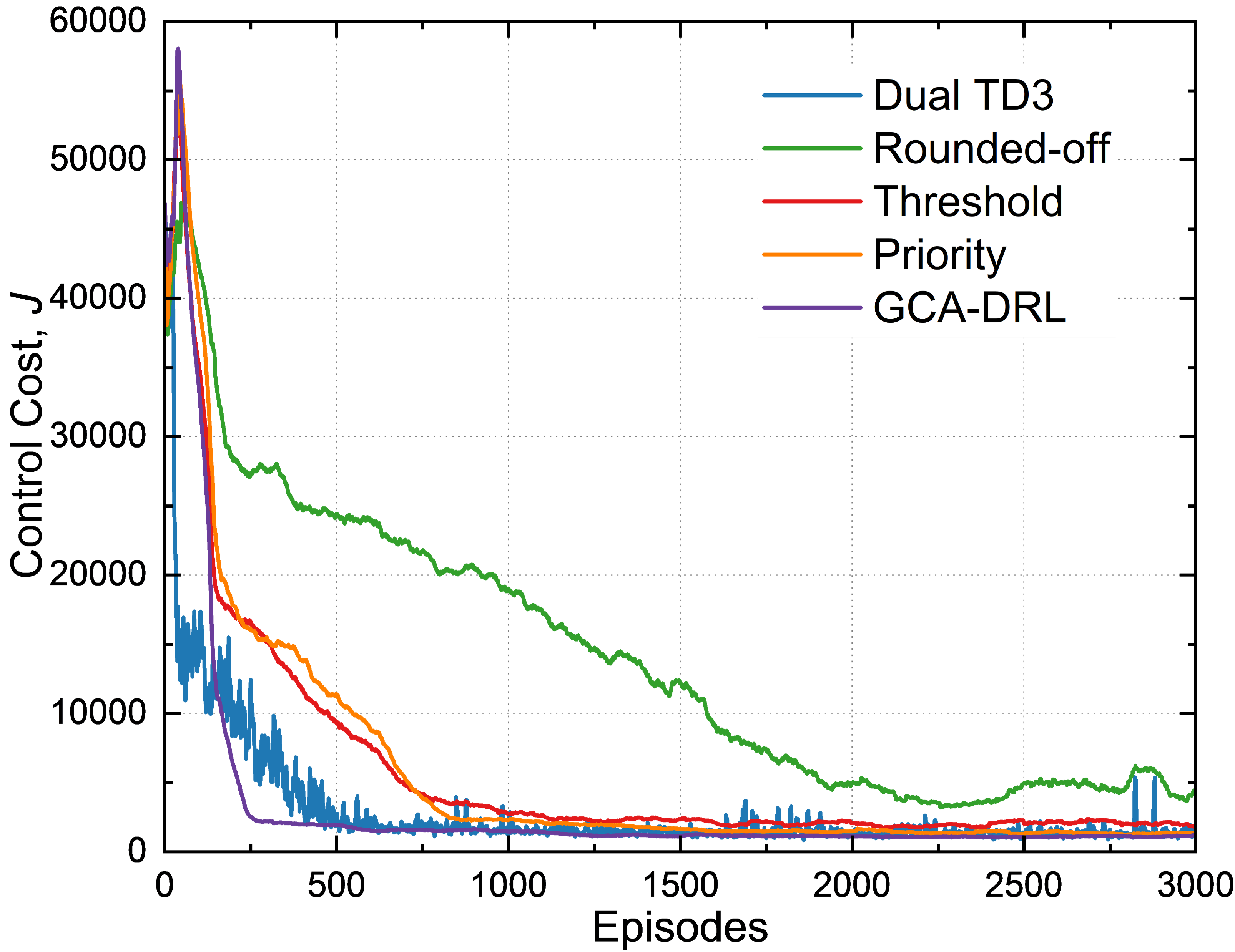}
	\vspace{-0.4cm}
	\caption{Control cost of five codesign methods using DRL during training.}
	\label{fig:TrainCompare}
	\vspace{0.0cm}
\end{figure}

Fig.~\ref{fig:TrainCompare} presents the training performance of five different codesign methods, including GCA-DRL, in terms of the control cost $J$ over 3000 training episodes. The GCA-DRL method demonstrates a clear advantage in both convergence speed and final performance. It not only converges faster but also achieves a significantly lower steady-state control cost compared to the other methods. By episode 500, GCA-DRL already exhibits a stable downward trend, whereas methods like dual-TD3 and the rounded-off method continue to show fluctuations and higher costs throughout the training period. The dual-TD3 method, in particular, shows the most variability and fails to converge to a similarly low control cost, with persistent oscillations around much higher values. The threshold-based and priority-based methods initially show relatively rapid convergence but plateau at higher control costs than GCA-DRL. These results highlight the superiority of GCA-DRL in terms of achieving both faster convergence and a lower overall control cost. This is attributed to the adaptive cascaded dual-branch actor architecture and the knowledge-assisted graph action embedding, which enable more effective codesign of communication and control policies.

Table~\ref{tab:Scalability} presents the comparison among the proposed GCA-DRL algorithm and several benchmark approaches in terms of the control cost across various system scales. The system scales are represented by different combinations of $L$, $N$, and $M$, which correspond to the number of wireless channels, actuators, and sensors, respectively. The GCA-DRL consistently achieves the lowest costs across all tested scales. All methods using DRL in Table~\ref{tab:Scalability} employ LN in \eqref{eq:LN}. Without LN, we can only train DRL with 10 devices sharing 5 channels. However, after introducing LN, we can train DRL with up to 40 devices sharing 20 channels. The results highlight the scalability of the proposed GCA-DRL algorithm that is effective for larger and more complex WNCSs.

\begin{table*}[t]
\centering
\footnotesize
\setlength\tabcolsep{3pt}
\caption{Comparison of the GCA-DRL and Benchmarks in the Control Cost $J$ with Different System Scales $(L,N,M)$ and $K=N$.}
\vspace{-0.3cm}
\begin{tabular}{m{2.1cm}m{3.4cm}m{2.2cm}m{1.0cm}m{1.0cm}m{1.1cm}m{1.3cm}m{1.3cm}m{1.3cm}m{1.3cm}}
\hline\hline
\textbf{Design approach} &\textbf{Scheduler} & \textbf{Controller} & $(4,5,6)$ & $(5,5,5)$ & $(8,10,12)$ & $(10,10,10)$  & $(12,15,18)$ & $(15,15,15)$  & $(20,20,20)$ \\ \hline\hline
\rowcolor[HTML]{EFEFEF} 
Traditional design &AoI-based greedy policy & Standard LQR & 1867237 & 2519.92  & 64589.39 & 8975058 & 641046.04 & 27256173 & 2422586.57\\ 
Traditional design &AoI-based greedy policy & Vanilla TD3 & 9504.11 & 1201.36 & 3003.25 &  5479.01 & 5340.31 & 10083.39 & 36124.73 \\ 
\rowcolor[HTML]{EFEFEF} 
Traditional design &TD3 (priority-based mapping) & Standard LQR & 26211.48 & 1801.76  & 2928.08 & 24025.68 & 19567.30 & 20406.11 & 114926.68\\ 

\hline

Codesign &TD3 (priority-based mapping) & Vanilla TD3 & 15898.98 & 1203.14 & 4604.62 &  25321.05 & 43510.40 & 31898.73 & 122493.84 \\ 
\rowcolor[HTML]{EFEFEF} 
Codesign & \multicolumn{2}{l}{Single TD3 (priority-based mapping and control)} & 9995.56 & 1255.82 & 3509.47 &  19077.32 & 55938.53 & 33965.72 & 96412.94 \\ 
\textbf{Codesign} & \multicolumn{2}{l}{\textbf{GCA-DRL (This work)}} & \textbf{8973.17} & \textbf{858.65} & \textbf{2214.60} & \textbf{3418.03} & \textbf{5049.50} &  \textbf{6239.07} &  \textbf{26274.48} \\ \hline\hline
\end{tabular}
\label{tab:Scalability}
\vspace{-0.6cm}
\end{table*}





\section{Conclusions}\label{sec:conclusion}
We have addressed the key challenges of communication-control codesign in large-scale WNCSs by developing a practical model and an advanced DRL approach. Our model captures real-world complexities, and by formulating the problem as an MDP, we have jointly optimized scheduling and control, effectively addressing their interdependencies. The proposed GCA-DRL algorithm efficiently manages the high-dimensional hybrid action space and improves training with layer normalization and partial policy gradient updates. Extensive simulations demonstrate that our approach significantly outperforms benchmark policies, proving its effectiveness for optimizing communication-control codesign in large-scale WNCSs. Future work will focus on extending this approach to nonlinear control systems and exploring the potential of distributed algorithm codesign rather than relying on centralized algorithms. Additionally, we aim to develop a more comprehensive framework for the codesign of communications, control, and computing. Another direction will involve investigating the most advanced integrated sensing and communications framework in WNCSs to enhance overall system performance.
    \balance
    
	\ifCLASSOPTIONcaptionsoff
	\newpage
	\fi

	\bibliographystyle{IEEEtran}

\begin{thebibliography}{10}
	\providecommand{\url}[1]{#1}
	\csname url@samestyle\endcsname
	\providecommand{\newblock}{\relax}
	\providecommand{\bibinfo}[2]{#2}
	\providecommand{\BIBentrySTDinterwordspacing}{\spaceskip=0pt\relax}
	\providecommand{\BIBentryALTinterwordstretchfactor}{4}
	\providecommand{\BIBentryALTinterwordspacing}{\spaceskip=\fontdimen2\font plus
		\BIBentryALTinterwordstretchfactor\fontdimen3\font minus
		\fontdimen4\font\relax}
	\providecommand{\BIBforeignlanguage}[2]{{%
			\expandafter\ifx\csname l@#1\endcsname\relax
			\typeout{** WARNING: IEEEtran.bst: No hyphenation pattern has been}%
			\typeout{** loaded for the language `#1'. Using the pattern for}%
			\typeout{** the default language instead.}%
			\else
			\language=\csname l@#1\endcsname
			\fi
			#2}}
	\providecommand{\BIBdecl}{\relax}
	\BIBdecl
	
	\bibitem{jin2023cloud}
	J.~Jin, K.~Yu, J.~Kua, N.~Zhang, Z.~Pang, and Q.-L. Han, ``Cloud-fog
	automation: Vision, enabling technologies, and future research directions,''
	\emph{IEEE Trans. Ind. Inf.}, vol.~20, no.~2, pp. 1039--1054, 2023.
	
	\bibitem{honghao2024cloud}
	H.~Lyu, J.~Yan, J.~Zhang, Z.~Pang, G.~Yang, and A.~J. Isaksson, ``Cloud--fog
	automation: Heterogenous applications over new-generation infrastructure of
	virtualized computing and converged networks,'' \emph{IEEE Ind. Elect. Mag.},
	2024.
	
	\bibitem{bhimavarapu2022unobtrusive}
	K.~Bhimavarapu, Z.~Pang, O.~Dobrijevic, and P.~Wiatr, ``Unobtrusive, accurate,
	and live measurements of network latency and reliability for time-critical
	internet of things,'' \emph{IEEE Inter. Things Magazine}, vol.~5, no.~3, pp.
	38--43, 2022.
	
	\bibitem{CoDesign}
	G.~Zhao, M.~A. Imran, Z.~Pang, Z.~Chen, and L.~Li, ``Toward real-time control
	in future wireless networks: Communication-control co-design,'' \emph{IEEE
		Commun. Mag.}, vol.~57, no.~2, pp. 138--144, 2019.
	
	\bibitem{Gatsis2015Control}
	K.~Gatsis, M.~Pajic, A.~Ribeiro, and G.~J. Pappas, ``Opportunistic control over
	shared wireless channels,'' \emph{IEEE Trans. Autom. Control}, vol.~60,
	no.~12, pp. 3140--3155, 2015.
	
	\bibitem{knorn2017optimal}
	S.~Knorn and S.~Dey, ``Optimal energy allocation for linear control with packet
	loss under energy harvesting constraints,'' \emph{Automatica}, vol.~77, pp.
	259--267, 2017.
	
	\bibitem{Chang2019Power}
	B.~Chang, L.~Zhang, L.~Li, G.~Zhao, and Z.~Chen, ``Optimizing resource
	allocation in {URLLC} for real-time wireless control systems,'' \emph{IEEE
		Trans. Veh. Technol.}, vol.~68, no.~9, pp. 8916--8927, 2019.
	
	\bibitem{lu2023jointly}
	J.~Lu and D.~E. Quevedo, ``A jointly optimal design of control and scheduling
	in networked systems under denial-of-service attacks,'' \emph{Automatica},
	vol. 148, p. 110774, 2023.
	
	\bibitem{leong2017event}
	A.~S. Leong, D.~E. Quevedo, T.~Tanaka, S.~Dey, and A.~Ahl{\'e}n, ``Event-based
	transmission scheduling and {LQG} control over a packet dropping link,''
	\emph{IFAC-PapersOnLine}, vol.~50, no.~1, pp. 8945--8950, 2017.
	
	\bibitem{Ji2022Edge}
	Z.~Ji, C.~Chen, J.~He, S.~Zhu, and X.~Guan, ``Edge sensing and control
	co-design for industrial cyber-physical systems: Observability guaranteed
	method,'' \emph{IEEE Trans. Cybern.}, vol.~52, no.~12, pp. 13\,350--13\,362,
	2022.
	
	\bibitem{Ji2023Edge}
	------, ``Learning-based edge sensing and control co-design for industrial
	cyber–physical system,'' \emph{IEEE Trans. Autom. Control}, vol.~20, no.~1,
	pp. 59--73, 2023.
	
	\bibitem{Tzoumas2021LQG}
	V.~Tzoumas, L.~Carlone, G.~J. Pappas, and A.~Jadbabaie, ``{LQG} control and
	sensing co-design,'' \emph{IEEE Trans. Autom. Control}, vol.~66, no.~4, pp.
	1468--1483, 2021.
	
	\bibitem{Wang2023TII}
	X.~Wang, J.~Zhang, C.~Chen, J.~He, Y.~Ma, and X.~Guan, ``Trust-{AoI}-aware
	codesign of scheduling and control for edge-enabled {IIoT} systems,''
	\emph{IEEE Trans. Ind. Inf.}, pp. 1--10, 2023.
	
	\bibitem{Girgis2021TCOMM}
	A.~M. Girgis, J.~Park, M.~Bennis, and M.~Debbah, ``Predictive control and
	communication co-design via two-way {Gaussian} process regression and
	{AoI}-aware scheduling,'' \emph{IEEE Trans. Comm.}, vol.~69, no.~10, pp.
	7077--7093, 2021.
	
	\bibitem{Wang2021Control}
	X.~Wang, C.~Chen, J.~He, S.~Zhu, and X.~Guan, ``{AoI}-aware control and
	communication co-design for industrial {IoT} systems,'' \emph{IEEE Internet
		Things J.}, vol.~8, no.~10, pp. 8464--8473, 2021.
	
	\bibitem{Ji2023Couple}
	Z.~Ji, C.~Chen, S.~Zhu, Y.~Ma, and X.~Guan, ``Intelligent edge sensing and
	control co-design for industrial cyber-physical system,'' \emph{IEEE Trans.
		Signal Inf. Process. Over Netw.}, vol.~9, pp. 175--189, 2023.
	
	\bibitem{lv2022impacts}
	H.~Lv, Z.~Pang, K.~Bhimavarapu, and G.~Yang, ``Impacts of wireless on robot
	control: the network hardware-in-the-loop simulation framework and real-life
	comparisons,'' \emph{IEEE Trans. Ind. Inf.}, vol.~19, no.~9, pp. 9255--9265,
	2022.
	
	\bibitem{lyu2024latency}
	H.~Lyu, Z.~Pang, A.~Bengtsson, S.~Nilsson, A.~J. Isaksson, and G.~Yang,
	``Latency-aware control for wireless cloud-fog automation: Framework and case
	study,'' \emph{IEEE Trans. Autom. Sci. Eng.}, 2024.
	
	\bibitem{Demirel2014TAC}
	B.~Demirel, Z.~Zou, P.~Soldati, and M.~Johansson, ``Modular design of jointly
	optimal controllers and forwarding policies for wireless control,''
	\emph{IEEE Trans. Autom. Control}, vol.~59, no.~12, pp. 3252--3265, 2014.
	
	\bibitem{Cao2023PAoL}
	J.~Cao, X.~Zhu, S.~Sun, P.~Popovski, S.~Feng, and Y.~Jiang, ``Age of loop for
	wireless networked control system in the finite blocklength regime: Average,
	variance and outage probability,'' \emph{IEEE Trans. Wirel. Commun.},
	vol.~22, no.~8, pp. 5306--5320, 2023.
	
	\bibitem{di2015co}
	G.~Di~Girolamo, A.~D'Innocenzo, and M.~D. Di~Benedetto, ``Co-design of
	controller and routing redundancy over a wireless network,''
	\emph{IFAC-PapersOnLine}, vol.~48, no.~22, pp. 100--105, 2015.
	
	\bibitem{Lu2023TII}
	X.~Lu, Q.~Xu, X.~Wang, M.~Lin, C.~Chen, Z.~Shi, and X.~Guan, ``Full-loop
	{AoI}-based joint design of control and deterministic transmission for
	industrial {CPS},'' \emph{IEEE Trans. Ind. Inf.}, vol.~19, no.~11, pp.
	10\,727--10\,738, 2023.
	
	\bibitem{9945199}
	M.~Lin, Q.~Xu, X.~Lu, J.~Zhang, and C.~Chen, ``Control and transmission
	co-design for industrial {CPS} integrated with time-sensitive networking,''
	in \emph{Proc. IEEE SMC}, 2022, pp. 229--234.
	
	\bibitem{Chen2020TII}
	C.~Chen, L.~Lyu, S.~Zhu, and X.~Guan, ``On-demand transmission for
	edge-assisted remote control in industrial network systems,'' \emph{IEEE
		Trans. Ind. Inf.}, vol.~16, no.~7, pp. 4842--4854, 2020.
	
	\bibitem{wildhagen2020scheduling}
	S.~Wildhagen and F.~Allg{\"o}wer, ``Scheduling and control over networks using
	mpc with time-varying terminal ingredients,'' in \emph{Proc. IEEE ACC}, 2020,
	pp. 1913--1918.
	
	\bibitem{9210732}
	P.~Li, Y.-B. Zhao, and Y.~Kang, ``Integrated channel-aware scheduling and
	packet-based predictive control for wireless cloud control systems,''
	\emph{IEEE Trans. Cybern.}, vol.~52, no.~5, pp. 2735--2749, 2022.
	
	\bibitem{Peters2016Codesign}
	E.~G.~W. Peters, D.~E. Quevedo, and M.~Fu, ``Controller and scheduler codesign
	for feedback control over {IEEE} 802.15.4 networks,'' \emph{IEEE Trans.
		Control Syst. Technol.}, vol.~24, no.~6, pp. 2016--2030, 2016.
	
	\bibitem{8796135}
	M.~Kögel, D.~Quevedo, and R.~Findeisen, ``Combined control and communication
	scheduling for constrained system using robust output feedback {MPC},'' in
	\emph{Proc. IEEE ECC}, 2019, pp. 1778--1783.
	
	\bibitem{yao2020contention}
	N.~Yao, M.~Malisoff, and F.~Zhang, ``Contention-resolving model predictive
	control for coupled control systems with a shared resource,''
	\emph{Automatica}, vol. 122, p. 109219, 2020.
	
	\bibitem{cui2020co}
	D.~Cui and H.~Li, ``Co-design of sampling pattern and control in self-triggered
	model predictive control for sampled-data systems,''
	\emph{IFAC-PapersOnLine}, vol.~53, no.~2, pp. 1795--1800, 2020.
	
	\bibitem{bahraini2022optimal}
	M.~Bahraini, M.~Zanon, A.~Colombo, and P.~Falcone, ``Optimal scheduling and
	control for constrained multi-agent networked control systems,'' \emph{Optim.
		Control Appl. Methods}, vol.~43, no.~1, pp. 23--43, 2022.
	
	\bibitem{redder2019deep}
	A.~Redder, A.~Ramaswamy, and D.~E. Quevedo, ``Deep reinforcement learning for
	scheduling in large-scale networked control systems,''
	\emph{IFAC-PapersOnLine}, vol.~52, no.~20, pp. 333--338, 2019.
	
	\bibitem{8619335}
	D.~Baumann, J.-J. Zhu, G.~Martius, and S.~Trimpe, ``Deep reinforcement learning
	for event-triggered control,'' in \emph{Proc. IEEE CDC}, 2018, pp. 943--950.
	
	\bibitem{funk2021learning}
	N.~Funk, D.~Baumann, V.~Berenz, and S.~Trimpe, ``Learning event-triggered
	control from data through joint optimization,'' \emph{IFAC J. Syst. Control},
	vol.~16, p. 100144, 2021.
	
	\bibitem{kesper2023toward}
	L.~Kesper, S.~Trimpe, and D.~Baumann, ``Toward multi-agent reinforcement
	learning for distributed event-triggered control,'' in \emph{Proc. Mach.
		Learn. Res.}, 2023, pp. 1072--1085.
	
	\bibitem{9561274}
	K.~Shibata, T.~Jimbo, and T.~Matsubara, ``Deep reinforcement learning of
	event-triggered communication and control for multi-agent cooperative
	transport,'' in \emph{Proc. IEEE ICRA}, 2021, pp. 8671--8677.
	
	\bibitem{shibata2023deep}
	------, ``Deep reinforcement learning of event-triggered communication and
	consensus-based control for distributed cooperative transport,'' \emph{Rob.
		Autom. Syst.}, vol. 159, 2023, {Art.} no. 104307.
	
	\bibitem{wan2023model}
	H.~Wan, H.~R. Karimi, X.~Luan, and F.~Liu, ``Model-free self-triggered control
	based on deep reinforcement learning for unknown nonlinear systems,''
	\emph{Int. J. Robust Nonlinear Control}, vol.~33, no.~3, pp. 2238--2250,
	2023.
	
	\bibitem{wan2023integrated}
	H.~Wan, H.~R. Karimi, X.~Luan, S.~He, and F.~Liu, ``Integrated learning
	self-triggered control for model-free continuous-time systems with
	convergence guarantees,'' \emph{Eng. Appl. Artif. Intell.}, vol. 123, 2023,
	{Art.} no. 106462.
	
	\bibitem{wang2023deep}
	R.~Wang and K.~Kashima, ``Deep reinforcement learning for continuous-time
	self-triggered control with experimental evaluation,'' \emph{Adv. Rob.},
	vol.~37, no.~16, pp. 1012--1024, 2023.
	
	\bibitem{Termehchi}
	A.~Termehchi and M.~Rasti, ``A learning approach for joint design of
	event-triggered control and power-efficient resource allocation,'' \emph{IEEE
		Trans. Veh. Technol.}, vol.~71, no.~6, pp. 6322--6334, 2022.
	
	\bibitem{Zhao2023IoTJ}
	Z.~Zhao, W.~Liu, D.~E. Quevedo, Y.~Li, and B.~Vucetic, ``Deep learning for
	wireless networked systems: A joint estimation-control-scheduling approach,''
	\emph{IEEE Internet Things J.}, vol.~11, no.~3, pp. 4535--4550, 2023.
	
	\bibitem{lima2022model}
	V.~Lima, M.~Eisen, K.~Gatsis, and A.~Ribeiro, ``Model-free design of control
	systems over wireless fading channels,'' \emph{Signal Process.}, vol. 197, p.
	108540, 2022.
	
	\bibitem{lei2016layer}
	J.~Lei~Ba, J.~R. Kiros, and G.~E. Hinton, ``Layer normalization,'' \emph{ArXiv
		e-prints}, pp. 1--14, 2016.
	
	\bibitem{10233620}
	Y.~Wang, S.~Wu, J.~Jiao, N.~Zhang, and Q.~Zhang, ``On the performance of joint
	resource allocation and control in industrial {IoT},'' in \emph{Proc.
		IEEE/CIC ICCC}, 2023, pp. 1--6.
	
	\bibitem{Sadeghi2008FSMC}
	P.~Sadeghi, R.~A. Kennedy, P.~B. Rapajic, and R.~Shams, ``Finite-state {Markov}
	modeling of fading channels - a survey of principles and applications,''
	\emph{IEEE Signal Process. Mag.}, vol.~25, no.~5, pp. 57--80, 2008.
	
	\bibitem{He2017FSMC}
	Y.~He \emph{et~al.}, ``Deep-reinforcement-learning-based optimization for
	cache-enabled opportunistic interference alignment wireless networks,''
	\emph{IEEE Trans. Veh. Technol.}, vol.~66, no.~11, pp. 10\,433--10\,445,
	2017.
	
	\bibitem{Liu2021Polyanskiy}
	W.~Liu, G.~Nair, Y.~Li, D.~Nesic, B.~Vucetic, and H.~V. Poor, ``On the latency,
	rate, and reliability tradeoff in wireless networked control systems for
	{IIoT},'' \emph{IEEE Internet Things J.}, vol.~8, no.~2, pp. 723--733, 2021.
	
	\bibitem{PangFBL}
	G.~Pang, W.~Liu, Y.~Li, and B.~Vucetic, ``{DRL}-based resource allocation in
	remote state estimation,'' \emph{IEEE Trans. Wirel. Commun.}, vol.~22, no.~7,
	pp. 4434--4448, 2022.
	
	\bibitem{fujimoto2018addressing}
	S.~Fujimoto, H.~Hoof, and D.~Meger, ``Addressing function approximation error
	in actor-critic methods,'' in \emph{Proc. ICML}, 2018, pp. 1587--1596.
	
	\bibitem{duff2001algorithms}
	I.~S. Duff and J.~Koster, ``On algorithms for permuting large entries to the
	diagonal of a sparse matrix,'' \emph{SIAM J. Matrix Anal. Appl.}, vol.~22,
	no.~4, pp. 973--996, 2001.
	
	\bibitem{Leong2020OMA}
	A.~S. Leong, A.~Ramaswamy, D.~E. Quevedo, H.~Karl, and L.~Shi, ``Deep
	reinforcement learning for wireless sensor scheduling in cyber-physical
	systems,'' \emph{Automatica}, vol. 113, 2020, {Art.} no. 108759.
	
	\bibitem{WNC1}
	L.~Schenato, B.~Sinopoli, M.~Franceschetti, K.~Poolla, and S.~S. Sastry,
	``Foundations of control and estimation over lossy networks,'' \emph{Proc.
		IEEE}, vol.~95, no.~1, pp. 163--187, 2007.
	
\end{thebibliography}

\end{document}